\documentclass[11pt, a4paper]{article}

\usepackage{amssymb}
\usepackage{latexsym}
\usepackage{psfig}
\usepackage{fullpage}

\newcommand{\beq}{\begin{equation}}
\newcommand{\eeq}{\end{equation}}
\newcommand{\ba}{\begin{array}}
\newcommand{\ea}{\end{array}}
\newcommand{\bea}{\begin{eqnarray}}
\newcommand{\eea}{\end{eqnarray}}
\newcommand{\bc}{\begin{center}}
\newcommand{\ec}{\end{center}}
\newcommand{\bt}{\begin{table}}
\newcommand{\et}{\end{table}}

\newcommand{\la}[1]{\label{#1}}

\newcommand{\bbox}{\rule{3mm}{3mm}}

\newcommand{\ds}{\displaystyle}
\newcommand{\no}{\noindent}
\newcommand{\pp}[2]{{\partial #1 \over \partial #2}}

\newcommand{\rf}[1]{(\ref{#1})}
\newcommand{\beqno}{\begin{displaymath}}
\newcommand{\eeqno}{\end{displaymath}}

\newcommand{\been}{\begin{enumerate}}
\newcommand{\een}{\end{enumerate}}

\newcommand{\sn}{{\rm sn}}
\newcommand{\cn}{{\rm cn}}
\newcommand{\dn}{{\rm dn}}

\newcounter{saveeqn}

\newcommand{\alpheqn}{\setcounter{saveeqn}{\value{equation}}
\stepcounter{saveeqn}\setcounter{equation}{0}
\renewcommand{\theequation}{\mbox{\arabic{saveeqn}\alph{equation}}}}

\newcommand{\resetalpheqn}{\setcounter{equation}{\value{saveeqn}}
\renewcommand{\theequation}{\arabic{equation}}}

\newtheorem{theo}{Theorem}

\newtheorem{example}{Proof}

\begin{document}

\title{Dynamics and Stability of Bose-Einstein Condensates:  the
  Nonlinear Schr\"odinger Equation with Periodic Potential}

\author{\bf Bernard Deconinck\thanks{Acknowledges support from the National 
  Science Foundation (DMS-0071568)}, Bela A. Frigyik, and J. Nathan 
  Kutz\thanks{Acknowledges support from the National Science Foundation 
  (DMS-0092682)},\\Department of Applied Mathematics,\\ 
  University of Washington,\\ Seattle, WA 98195-2420, USA} 

\maketitle
\begin{abstract}
  The cubic nonlinear Schr\"odinger equation with a lattice potential is used
  to model a periodic dilute gas Bose-Einstein condensate.  Both two- and
  three-dimensional condensates are considered, for atomic species with either
  repulsive or attractive interactions.  A family of exact solutions and
  corresponding potential is presented in terms of elliptic functions.  The
  dynamical stability of these exact solutions is examined using both
  analytical and numerical methods.  For condensates with repulsive atomic
  interactions, all stable, trivial-phase solutions are off-set from the zero
  level.  For condensates with attractive atomic interactions, no stable
  solutions are found, in contrast to the one-dimensional case \cite{becpre2}.
\end{abstract}


\section{Introduction}

The cubic nonlinear Schr\"odinger equation with repulsive or attractive
nonlinearity and an external potential models the mean-field dynamics of a
dilute-gas Bose Einstein condensate (BEC) \cite{gross,pitaevskii}. In this
context, the equation is also known as the Gross-Pitaevskii equation.  BECs
are of interest in the theoretical and experimental physics community since
they are examples of macroscopic quantum phenomena which manifest
phase coherence~\cite{anderson1,chio,hagley1,hagley2}.  BECs
have possible applications in such different areas as quantum logic
\cite{brennen}, matter-wave diffraction \cite{diffraction}, and matter-wave
transport \cite{transport}.

After rescaling of the physical variables, the governing equation is 
\beq
\label{eqn:nls}
 i \pp{}{t}\psi(\vec{x},t)= -\frac{1}{2}\Delta\psi(\vec{x},t) + 
 \alpha \left|\psi(\vec{x},t)\right|^2 \psi(\vec{x},t) 
        + V(\vec{x}) \psi(\vec{x},t),
\eeq
where $\Delta$ denotes the Laplacian operator, $\psi(\vec{x},t)$ is the
macroscopic wave function of the condensate in one, two or three dimensions,
with $\vec{x}$ defined as $x$, $(x,y)$ or $(x,y,z)$ respectively.  The
real-valued function $V(\vec{x})$ is an experimentally generated macroscopic
potential. The parameter $\alpha$ determines whether (\ref{eqn:nls}) is
repulsive ($\alpha=1$, defocusing nonlinearity), or attractive, ($\alpha=-1$,
focusing nonlinearity).  The dimensionless variables in (\ref{eqn:nls}) are
related to the physical variables by $t\rightarrow (4\pi \hbar |a|
N/M\Omega)t$, $\vec{x}\rightarrow (\Omega/4\pi |a| N)^{1/2} \vec{x}$,
$\psi\rightarrow (\hbar\Omega)^{-1/2}\psi$, and $V(\vec{x})\rightarrow
(M\Omega/4\pi |a| N) V(\vec{x})$.  Here $N$ is the number of atoms in the
volume $\Omega$ containing all atoms (condensed and uncondensed) in the
the experiment 
(typically on the order of ($10\mu$m)$^3$
\cite{rubidium1}), $M$ is the mass of a single atom, and $a$ is the $s$-wave
scattering length for collisions between atoms.  Depending on the atomic
species, $a$ can be either positive or negative so that both signs of
$\alpha={\rm sign}(a)$ are relevant for BEC applications.  In Table
\ref{table}, various physical quantities for BEC experiments are displayed.
Note that the $s$-wave scattering lengths for $^7$Li and
$^85$Rb are negative, so these
condensates are attractive, whereas the other experiments listed use repulsive
condensates.

%
\begin{table}[t]
\begin{center}
  \begin{tabular}{ccccccc}
\hline\hline
 atom & $a$ [nm] 
    & $T_c$ [$\mu$K] & $N_c$ [$\times 10^{5}$] &  condensate lifetime  
       [s]  & $\Delta t$ & reference \\ \hline\hline 
  $^1$H     & 0.072 & 50   & 200,000 & 5         & 3,200 & \cite{hydrogen} \\
  $^{23}$Na & 2.75  &  2   &      7  & 1         & 6,700 & \cite{sodium1} \\
  $^{87}$Rb & 5.77  & 0.17 &   0.2   & 15        & 2     & \cite{rubidium1} \\
  $^{87}$Rb & 5.77  & 0.43 &    15   & $<$25$^*$  & 74    & \cite{rubidium2} \\
            &       &      &         &            &       &    \\
  $^7$Li    & -1.44     & 0.30   & 1 & 0.1$^\dag$  & 0.15  & \cite{lithium} \\
  $^{85}$Rb   & -0.9$^\ddag$  
    & 0.006  &     15  &  0.012    & 0.0007  & \cite{rubidium85} \\
 \hline\hline
  \end{tabular}
\caption{ \la{table}  Physical parameters for various BEC experiments along
  with the lifetime of the condensate and the elapsed scaled time $\Delta t$ 
  for \rf{eqn:nls}.  Note that both $^7$Li and $^{85}$Rb 
  have a negative $s$-wave scattering
  length which imply an attractive condensate.  The last column of the
  table gives the reference for the corresponding experiment.  For the
  data in the table: 
  $*$ denotes the total time of the experiment, 
  $\dag$ is estimated from the peak width of Fig.~3 of \cite{lithium},
  $\ddag$ is calculated using the experimental results of \cite{rubidium85}.
}
\end{center}
\end{table}
%

Current BEC experiments require a confining potential in order to trap the
condensate.  For theoretical convenience, this confinement is often assumed to
be harmonic.  Once confined, additional electromagnetic potentials may be
imposed on the BEC.  In particular, there has been recent interest in
confinement of repulsive BECs using standing light waves which results in a
sinusoidal potential in the Nonlinear Schr\"odinger equation~(\ref{eqn:nls}).
Although experiments focus on the quasi-one-dimensional regime,
quasi-two-dimensional and three-dimensional lattice potentials have been
suggested~\cite{jaksch}.  The interaction of the BECs with such potentials is
of great practical and theoretical interest.  The quasi-one-dimensional
regime~\cite{becpre1,petrov1D} applies when the longitudinal dimension of the
condensate is far greater than the transverse dimensions, which are themselves
of the order of the healing length of the condensate.  Likewise, the
quasi-two-dimensional regime~\cite{petrov2D} is significant when two
dimensions of the condensate are far greater than the third dimension, which
is itself of the order of the healing length of the condensate.  The healing
length is the width of the boundary region over which the probability density
of the condensate drops to zero.  Thus these quasi-one- or two-dimensional
approximations allow for simplification of (\ref{eqn:nls}).  In reality,
however, all condensates are three-dimensional and in many cases no reduction
in dimension is justifiable.

\begin{figure}[t]
\centerline{\psfig{figure=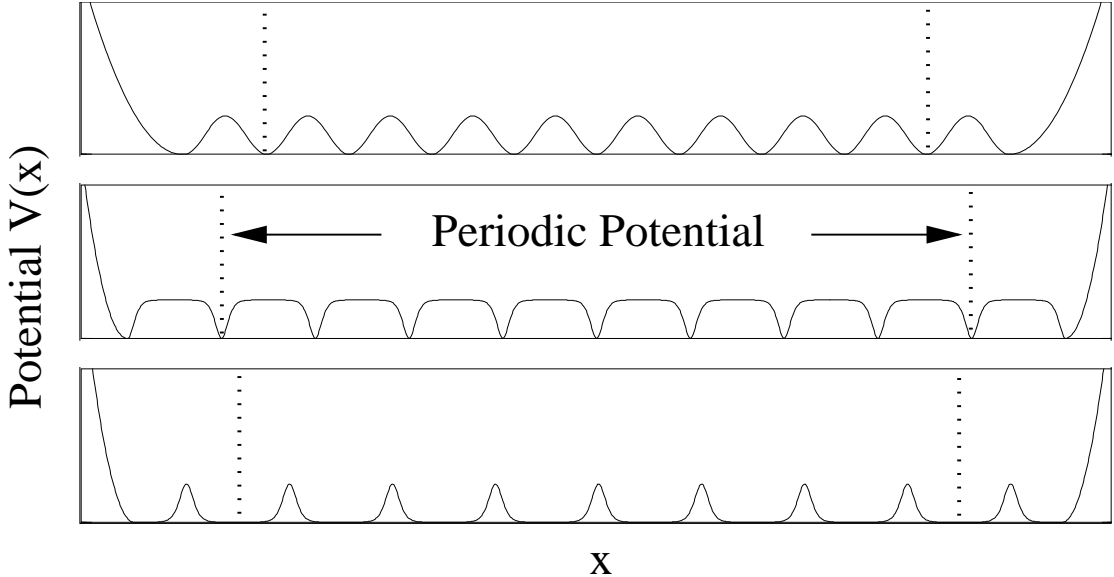,width=150mm,silent=}}
\begin{center}
\caption{ \label{fig:pots1D} External potential $V(x)$ which
exhibits both a periodic and a confining structure.  The
periodic structure is given by $sn^2(x,k)$ which reduces
to a standing light wave (as $k\rightarrow 0$) in the top figure. Other
possibilities of the potential \rf{eqn:pot1} include well-separated troughs or
peaks. 
In this paper, we consider the periodic region away from the edge of the confining
experimental potential.   Typically this region extends over tens 
of periods~\cite{kasevich}.
}
\end{center}
\end{figure}

In one dimension, the potential generated in experiments is a standing light
wave~\cite{bongs} and $V(x)=-V_0~\sin^2(m x)=(V_0~\cos(2 m x)-V_0)/2$.  The
same potential was considered in \cite{berg,choi}.  Typically, the condensate
is distributed over tens of periods of the standing light
wave~\cite{kasevich}.  In \cite{becpre2,becprl1,becpre1}, a number of exact
solutions of \rf{eqn:nls} with this potential was constructed and their
stability was investigated.  Additionally, a more general potential was
considered:
\beq\la{eqn:pot1}
V(x)=-V_0~\sn^2(m x,k),
\eeq
where $\sn(x,k)$ denotes the Jacobian elliptic sine function, with elliptic
modulus $k$. It is periodic with period $4K(k)=4 \int_{0}^{\pi/2}
d\alpha/\sqrt{1-k^2 \sin^2 \alpha}$.  This family of potentials contains the
standing light wave potential as a special case: $\lim_{k\rightarrow
0}\sn(x,k)=\sin(x)$. This more general potential is considered because it
allows the inclusion of different regimes of BEC dynamics.  In particular, for
values of $k$ up to $k=0.9$, the shape of the potential is virtually
indistinguishable from a sinusoidal one, but for values of $k$ close to $1$,
$i.e.,$ $k>0.999$, \rf{eqn:pot1} gives periodic potentials with well-separated
peaks or troughs.  Figure \ref{fig:pots1D} illustrates the different regimes
of the periodic potential \rf{eqn:pot1}.  The figure also indicates the
presence of a confining potential which is applied in all current experiments.

Due to technical complications, there are currently no experiments where a BEC
is trapped in a higher-than-one-dimensional periodic potential.  However, the
interest in the applications mentioned above strongly suggests that these
experiments will eventually take place.  Already, theoretical investigations
of BECs in multidimensional lattice potentials suggest the realization of such
experiments \cite{jaksch}.  Motivated by the previously referenced
developments in BECs, we consider (\ref{eqn:nls}) with repulsive and
attractive nonlinearity and periodic potential in two and three dimensions.
Note that in \cite{pla}, the repulsive condensate in two-dimensions was
investigated. 

A judicious choice of the potential allows the construction of a large class
of exact solutions, as in one dimension~\cite{becpre2,becprl1,becpre1,pla}.
The family of potentials considered is
\beq\la{eqn:pot}
V(\vec{x})=-\prod_{i=1}^d \left(A_i \sn^2(m_i x_i,k_i)+B_i\right)+
\sum_{i=1}^d m_i^2 k_i^2 \sn^2(m_i x_i,k_i),
\eeq
where $d$ is the number of dimensions: $d=2$ or $d=3$, and $x_1=x$, $x_2=y$, and
$x_3=z$.  Here  $A_i$, $B_i$ and
$m_i$, $i=1,\ldots, d$ are real constants. The elliptic moduli $k_i$ are in the
interval $[0,1]$.  The first term is a straightforward generalization of the
one-dimensional potential (\ref{eqn:pot1}), with an additive constant.  The
other terms facilitate the construction of exact solutions. 

\begin{figure}[tb]
\centerline{\psfig{figure=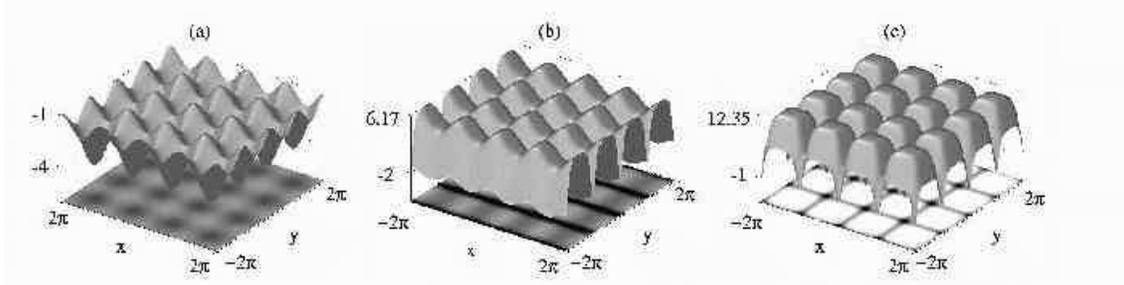,width=150mm,silent=}}
\begin{center}
\caption{ \label{fig:pots} Various lattice potentials. For all figures,
$A_1=A_2=B_1=B_2=1$. For (a), $k_1=k_2=0$, $m_1=m_2=1$; for (b), $k_1=0.999$,
$k_2=0$, $m_1=2K(0.999)/\pi$, $m_2=1$. Finally, for (c) $k_1=k_2=0.999$ and
$m_1=m_2=2K(0.999)/\pi$.}
\end{center}
\end{figure}

\begin{figure}[thb]
\centerline{\psfig{figure=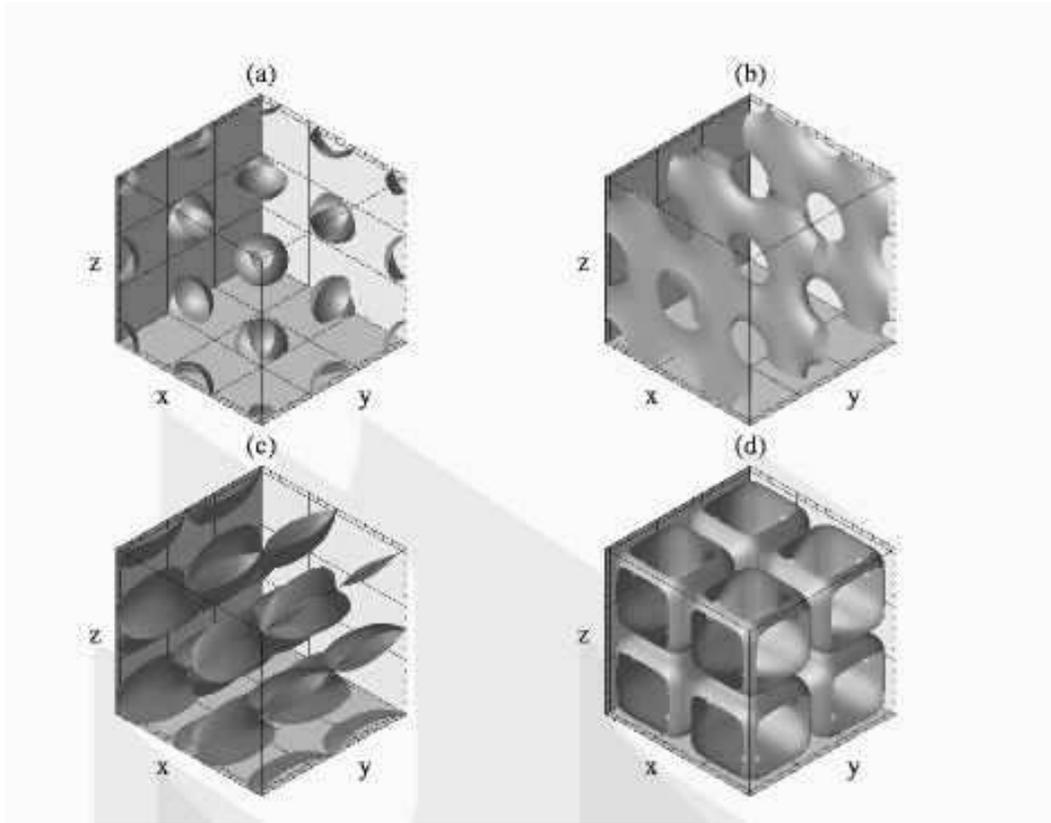,width=140mm,silent=}}
\begin{center}
\caption{ \label{fig:pots3D} Various 3-D lattice potentials. For all figures,
$A_1=A_2=B_1=B_2=-1$. 
For (a), $k_1=k_2=k_3=0$ and $m_1=m_2=m_3=1$; 
for (b), $k_1=k_2=0$, $k_3=0.999$, $m_1=m_2=1$, $m_3=2K(0.999)/\pi$,
for (c), $k_1=0$, $k_2=k_3=0.999$, $m_1=1$, $m_2=m_3=2K(0.999)/\pi$,
for (d), $k_1=k_2=k_3=0.999$, $m_1=m_2=m_3=2K(0.999)/\pi$. }
\end{center}
\end{figure}

Note that in the important physical case of a trigonometric potential $k_i
\rightarrow 0$, $i=1, \ldots, d$, the potential reduces to a product of
standing light wave potentials. This generalization of the one-dimensional
case is completely analogous with the standard approach of higher-dimensional
Kronig-Penney potentials \cite{kronig}.  For non-trigonometric potentials, the
exact expression \rf{eqn:pot} for the potential is necessary to allow the
construction of exact solutions. Nevertheless, it is the qualitative features
of the potential, $i.e.,$ its periodicity and amplitude, that are most
important for the stability properties of the solution. As numerical and
analytical results throughout this paper demonstrate, the behavior of a
solution in a lattice potential is largely independent of the quantitative
features of the potential, just as in the one-dimensional case
\cite{becpre2,becprl1,becpre1}.  Figures \ref{fig:pots} and \ref{fig:pots3D}
display the potential \rf{eqn:pot} for various values of its parameters in two
and three dimensions respectively.

Because of the present lack of experiments in two or three dimensions, this
paper can either be read as considering \rf{eqn:nls} with a periodic potential
as a mathematical problem in its own right, or as a collection of predictions
from the mathematical mean-field model concerning the dynamics and stability 
of BECs in higher-dimensional periodic potentials.   The paper is arranged
as follows:  exact solutions of \rf{eqn:nls} with periodic potential 
\rf{eqn:pot} are constructed.  A linear stability analysis is presented for
these exact solutions.  However, since in many cases the analysis is
inconclusive, a complete determination of the stability properties of all
our exact solutions is obtained by numerical simulation.  We conclude the
paper with a brief summary of the results and their implications for
BEC dynamics.

\section{Exact solutions}\la{sec:exact}

For typical potentials of \rf{eqn:nls}, the construction of exact solutions is
not obvious and usually not possible. For the potentials \rf{eqn:pot1} and
\rf{eqn:pot}, it is possible to find families of exact stationary solutions.
We look for solutions whose spatial dependence is separated:

\beq\la{eqn:sep}
\psi(\vec{x},t)=e^{-i \omega t}\prod_{i=1}^d \phi_i(x_i).  
\eeq

\no Here $d=2$ or $d=3$. The derivation for $d=1$ is similar, but slightly
easier.  Inserting the ansatz \rf{eqn:sep} into \rf{eqn:nls} gives 
\beq\la{eqn:separate}
\omega=-\frac{1}{2}\sum_{i=1}^d \frac{\phi_i''}{\phi_i}+
\alpha \prod_{i=1}^d |\phi_i|^2+V(\vec{x}),
\eeq

\no where prime denotes differentiation with respect to the argument.
In order to separate this equation, the potential is written as 

\beq\la{eqn:potguess}
V(\vec{x})=-\alpha \prod_{i=1}^d |\phi_i|^2+\sum_{i=1}^d W_i(x_i),
\eeq

\no which results in the $d$ separated equations

\beq\la{eqn:separated}
\omega_i \phi_i=-\frac{1}{2}\phi_i''+W_i(x_i) \phi_i,
\eeq

\no with $\sum_{i=1}^d \omega_i=\omega$. Using an amplitude-phase decomposition

\beq\la{eqn:ampphase}
\phi_i(x_i)=r_i(x_i) e^{i \theta_i(x_i)},
\eeq

\no eliminating the overall exponential factor, and equating the imaginary part
to zero, gives

\beq\la{eqn:phase}
\theta_i(x_i)=c_i \int_{0}^{x_i} \frac{d w}{r_i^2(w)},
\eeq

\no where $c_i$ is an integration constant. Multiplying the remaining equation
by $r_i'$, results in

\beq\la{eqn:integro}
\omega_{i} r_i^2=-\frac{1}{2} (r_i')^2-\frac{c_i^2}{2 r_i^2}+2\int W_i(x_i)
r_i(x_i) r_i'(x_i) dx_i.
\eeq

\no This equation is rewritten using the new function $S_i=r_i^2$. Combining this
with a substitution on the integral term finally gives

\beq\la{eqn:finally}
(S_i')^2=-8 \omega_i S_i^2-4 c_i^2+8 S_i\int \tilde{W}_i(S_i)d S_i.
\eeq

\no  Provided the right-hand side of this equation is a polynomial in $S_1$ of 
degree 3 or 4,  the equation is solvable in terms of elliptic functions \cite{ww}.
This imposes conditions on the potential for this construction to work: 
$\tilde{W}_i$ as a
function of $S_i$ is a polynomial of degree at most 2 \cite{ww}.   
Using this result, one finds

\beq\la{eqn:density}
r_i^2=A_i \sn^2(m_i x_i, k_i)+B_i,
\eeq

\no which gives rise to the potentials \rf{eqn:pot}. The parameters $\omega$,
$A_i$, $B_i$, $c_i$ are constrained by

\bea\la{eqn:omegai}
\omega&=& \sum_{i=1}^{d} \omega_i =  \sum_{i=1}^{d} \left(
  \frac{1}{2} m_i^2 (1+k_i^2)+ m_i^2 k_i^2
\frac{B_i}{A_i} \right), \\\la{eqn:ci}
c_i^2&=&m_i^2\frac{B_i}{A_i}(A_i+B_i)(A_i+k_i^2 B_i).
\eea

\no For $d=2$, this results in a one-parameter family of solutions: specifying
the potential fixes the parameters $A_1 A_2$, $B_1/A_1$, and $B_2/A_2$. Thus
$A_1/A_2$ is a free parameter. Similarly, for $d=3$, the cross ratio
$(A_1:A_2:A_3)$ is free, resulting in a two-parameter family of solutions. The
existence of these solutions requires $r_i^2\geq 0$ and $c_i^2\geq 0$. This
imposes additional constraints on the parameters: $A_i \geq 0$, $B_i\geq 0$, or
$B_i\geq 0$, $-A_i\leq B_i \leq -A_i/k_i^2$, $i=1, \ldots, d$. 

Note that
all our solutions satisfy
\bea\la{eqn:floquet} 
\psi(\vec{x}+2K(k_i)\vec{e_i},t)=\psi(\vec{x},t) e^{i \gamma_i},~~~~ \gamma_i= 
     c_i \int_{0}^{2K(k_i)} \frac{dw}{r_i^2(w)}, 
\eea
\no where $\vec{e_i}$ is the unit vector in the $x_i$ direction. 
Thus the solutions are nonlinear generalizations of the
Bloch states of solid state theory~\cite{ashcroft}. Note, however,  that any
such analogy is flawed in principle: the reasoning leading to Bloch
functions in solid state theory is based on methods for linear differential
equations, whereas the governing equation
\rf{eqn:nls} is nonlinear. Similarly flawed is any interpretation of $\omega$
as an eigenvalue or a separation constant.  However, for small amplitude
solutions for which the linear terms dominate the nonlinearity, 
the intuition gained from solid state band-gap theory can be useful in interpreting
the weakly nonlinear results.

From (\ref{eqn:density}), it follows that the amplitude $r_i(x_i)$ is 
periodic in $x_i$ with
period $2K(k_i)/m_i$.  However, generally the phase $\theta_i(x_i+2K(k_i)/m_i)\neq
\theta_i(x_i)+2 n \pi$, for some integer $n$.  Thus, the solution
$\psi(\vec{x},t)$
is  usually not periodic in any of the $x_i$, see \rf{eqn:floquet}. 
Imposing periodicity in the
$x_i$-direction  requires a quantization of the phase $\theta_i(x_i)$.   It is
unclear if such a quantization in all directions is possible, since not enough
free parameters are available to satisfy the number of quantization conditions.
There are two special cases in which phase quantization is not a concern. The
first case results in trivial-phase solutions, for which $c_i=0$. The
second case is the trigonometric limit, in which $k_i=0$, $i=1, \ldots, d$.

The solution has trivial phase in the $x_i$-direction if $c_i$ is zero.
There are three possibilities: 
\alpheqn
\bea \la{eqn:trivphasea}
B_i&=&0: ~~~\,~~~~~~~ r_i=\sqrt{A_i}~\sn(m_i x,k_i),\\ \la{eqn:trivphaseb}
B_i&=&-A_i: ~~\,~~~~ r_i=\sqrt{-A_i}~\cn(m_i x,k_i),\\ \la{eqn:trivphasec}
B_i&=&-A_i/k_i^2: ~~ r_i=\sqrt{-A_i/k_i^2}~\dn(m_i x,k_i),
\eea
\resetalpheqn
\no where $\cn(m_i x_i,k_i)$ is the Jacobian elliptic cosine function, and $\dn(m_i
x_i, k_i)$ denotes the third Jacobian elliptic function. 

The solution is trigonometric in the $i$-direction if $k_i$ is zero. Then
\beq\la{eqn:trig}
r_i^2(x)=A_i\sin^2(x_i)+B_i, ~~
\tan(\theta_i)=\sqrt{1+\frac{A_i}{B_i}}\tan(m_i x_i),
\eeq
and phase quantization is satisfied. 
Notice that it is possible for the solution to have trivial phase in one
direction and be trigonometric in the other.  In contrast,
as $k_i \rightarrow 1$ it is possible to obtain solitary wave solutions. These
are discussed in \cite{becpre2,becpre1}. More details on the trigonometric
solutions and the trivial phase solutions is also given there.


\section{Linear stability}\la{sec:stability}

The linear stability analysis of the solutions found in the previous sections
is completely analogous to the one-dimensional case, discussed in
\cite{becpre2,becpre1}.  The dimensionality of the solutions only requires 
minor alterations. 

Consider an infinitesimally small perturbation of an exact solution
\beq\la{eq:perturb}
\psi(\vec{x},t)=\left(r(\vec{x})+\epsilon \varphi(\vec{x},t)\right) \exp
\left(i \theta(\vec{x})-i \omega t\right),
\eeq
with $r(\vec{x})=\prod_{i=1}^d r_i(x_i)$ and $\theta(\vec{x})=\sum_{i=1}^d
\theta_i(x_i)$ and $r_i(x_i)$, $\theta_i(x_i)$ defined in \rf{eqn:density} and
\rf{eqn:phase} respectively. 
Substituting this in \rf{eqn:nls} and linearizing in $\epsilon$ gives
\beq\la{eqn:linearize}
i \pp{\varphi}{t}+\omega\varphi=-\frac{1}{2} \Delta \varphi+
\alpha r^2 (\varphi^*+2
\varphi)+V(\vec{x}) \varphi+\frac{1}{2} (\nabla \theta)^2 \varphi-\frac{i}{2} 
\varphi 
\Delta \theta-i \nabla \theta \cdot \nabla \varphi, 
\eeq
\no where $\varphi^*$ is the complex conjugate of $\varphi$,
$\nabla \theta \cdot \nabla \varphi=\sum_{i=1}^d 
\pp{\theta}{x_i}\pp{\varphi}{x_i}$,
and $(\nabla \theta)^2=\nabla \theta \cdot \nabla \theta$. Decomposing $\varphi$
into its real and imaginary parts $\varphi(\vec{x},t)=u(\vec{x},t)+i 
v(\vec{x},t)$, \rf{eqn:linearize} becomes 
\beq\la{uv}
\pp{}{t}\left(\ba{c}u\\v\ea\right)=J\left(\ba{rr}
L_+ & -S\\
S & L_-
\ea\right)
\left(\ba{c}u\\v\ea\right),
\eeq
\no with $J=\left(\ba{rr}0 & 1\\ -1 & 0\ea\right)$ and
\alpheqn
\bea\la{eqn:lplus}
L_+&=&-\frac{1}{2}\Delta+3\alpha
r^2+V(\vec{x})-\omega+\frac{1}{2}\left(\nabla \theta\right)^2,\\\la{eqn:lminus}
L_-&=&-\frac{1}{2}\Delta+\alpha
r^2+V(\vec{x})-\omega+\frac{1}{2}\left(\nabla \theta\right)^2,\\
S&=&-\nabla \theta \cdot \nabla -\frac{1}{2}\Delta \theta.
\eea
\resetalpheqn

For solutions with non-trivial phase, the analysis of this linear system is
currently an open problem.  We restrict ourselves to the stability analysis of
trivial-phase solutions: $\nabla \theta=0$. Separating the time dependence from
the spatial variations, 
\beq\la{eqn:separation}
\left(\ba{c}u\\v\ea\right)=\left(\ba{c}U\\V\ea\right)e^{\lambda} t,
\eeq
we obtain for trivial-phase solutions
\beq\la{system}
\lambda\left(\ba{c}U\\V\ea\right)=J\left(\ba{rr}
L_+ & 0\\
0 & L_-
\ea\right)
\left(\ba{c}U\\V\ea\right),
\eeq
\no with $L_+=-\frac{1}{2}\Delta+3\alpha r^2+V(\vec{x})-\omega$ and 
$L_-=-\frac{1}{2}\Delta+\alpha r^2+V(\vec{x})-\omega$. 
Note that the perturbations on the trivial-phase solutions considered do
not necessarily have a trivial-phase profile, since Im$(\varphi)=v$ is not
necessarily zero. 

Adding the potential $V(\vec{x})$ to the Nonlinear Schr\"odinger equation
destroys three of its four continuous symmetries. However, equation \rf{eqn:nls}
is still phase invariant: the transformation $\psi(\vec{x},t)\rightarrow e^{i
\gamma} \psi(\vec{x},t)$, with $\gamma$ a real constant, 
leaves \rf{eqn:nls} unchanged. Using Noether's theorem,
\beq\la{eqn:noether}
L_- r(\vec{x})=0. 
\eeq
Thus, $\lambda=0$ is in the spectrum of the Schr\"odinger operator $L_-$: $0 \in
\sigma(L_-)$. Since both $L_+$ and $L_-$ are self-adjoint operators, their
spectrum is contained on the real axis. Furthermore, since $-\frac{1}{2}\Delta$
is a positive operator, and $r(\vec{x})$, $V(\vec{x})$ are bounded, these
spectra are contained in some semi-infinite interval: 
\beq\la{eqn:contained}
\sigma(L_+)\in [\lambda_+,\infty), ~~~
\sigma(L_-)\in [\lambda_-,\infty), 
\eeq 
\no with $\lambda_\pm=\inf_{||\eta||=1}\left<\eta|L_{\pm}|\eta\right>$, 
$||\eta||^2=\int_{\Omega}|\eta|^2 d\vec{x}$, $\Omega=\left\{(x,y)\in 
[0,2K(k_1)]\times[0,2K(k_2)]\right\}$ for $d=2$, and $\Omega=\left\{(x,y,z)\in 
[0,2K(k_1)]\times[0,2K(k_2)]\times[0,2K(k_3)]\right\}$ for $d=3$.  Note that 
since $0 \in \sigma(L_-)$, $\lambda_-\leq 0$.

\begin{theo}\la{theo1}
For condensates with repulsive atomic interactions ($\alpha=1$)
an exact trivial-phase solution such that $r(\vec{x})>0$ is linearly
stable.
\end{theo}

\no {\bf Proof}~~~
In this case $\alpha=1$ and $L_+=L_-+2r^2$. Thus, $\lambda_+>\lambda_-$. 
If $r(\vec{x})>0$, then it is the ground state of $L_-$ \cite{ch1} and 
$\lambda_-=0$, $\lambda_+>0$. Note that for one-dimensional Schr\"odinger
operators, the ground state is unique. This is not necessarily true for
the higher-dimensional case we are considering.  
Since $L_+$ is a positive operator, we can construct $L_+^{1/2}$.
Consider 
\beq\la{eqn:h1}
H=L_+^{1/2} L_- L_+^{1/2}.
\eeq
Let $\xi=L_+^{1/2} U$, then
\beq\la{eqn:new1}
(H+\lambda^2)\xi=0.
\eeq
\no The operator $H$ is also self-adjoint. Let $\mu_0$ be its smallest
eigenvalue, then
\begin{eqnarray*}
\mu_0&=&\inf_{||\eta||=1}\left<\eta|H|\eta\right>\\
&=&\inf_{||\eta||=1}\left<\eta\left|L_+^{1/2}L_-L_+^{1/2}\right|\eta\right>\\
&=&\inf_{||\eta||=1}\left<L_+^{1/2}\eta\Big|L_-\Big|L_+^{1/2}\eta\right>,
\end{eqnarray*}
which implies $\mu_0\geq 0$.  Note that by letting
$\eta=L_+^{-1/2}r/||L_+^{-1/2}r||$, 
$\left<L_+^{1/2}\eta\Big|L_-\Big|L_+^{1/2}\eta\right>=0$ so that
$\mu_0=0$.
Thus $\lambda^2\leq0$,
and $\lambda$ is imaginary or zero. \hspace*{\fill}$\bbox$\\

\no {\bf Remark}~~~
As in \cite{becpre1},
we have proven the nonexistence of exponentially unstable eigenvalues. 
However, algebraic instabilities of the zero or other modes are not 
ruled out.

\begin{theo}\la{theo2}
For condensates with attractive atomic interactions ($\alpha=-1$)
an exact trivial-phase solution such that $r(\vec{x})>0$ is linearly
unstable.
\end{theo}

\no {\bf Proof}~~~
In this case $\alpha=-1$ and $L_-=L_++2r^2$. Thus, $\lambda_->\lambda_+$. 
If $r(\vec{x})>0$, then, as above, 
it is the ground state of $L_-$ \cite{ch1} and 
$\lambda_-=0$, $\lambda_+<0$.   
Since $L_-$ is a positive operator, we can construct $L_-^{1/2}$.
Consider 
\beq\la{eqn:h2}
H=L_-^{1/2} L_+ L_-^{1/2}.
\eeq
Let $\xi=L_-^{1/2} U$, then
\beq\la{eqn:new2}
(H+\lambda^2)\xi=0.
\eeq
\no The operator $H$ is self-adjoint. Again, let $\mu_0$ be its smallest
eigenvalue, then
\begin{eqnarray*}
\mu_0&=&\inf_{||\eta||=1}\left<\eta|H|\eta\right>\\
&=&\inf_{||\eta||=1}\left<\eta\left|L_-^{1/2}L_+L_-^{1/2}\right|\eta\right>\\
&=&\inf_{||\eta||=1}\left<L_-^{1/2}\eta\Big|L_+\Big|L_-^{1/2}\eta\right>\\
&<& 0.
\end{eqnarray*}
Thus there is at least one $\lambda^2$ such that $\lambda^2>0$, resulting in a
pair of opposite, real $\lambda$'s. 
\hspace*{\fill}$\bbox$\\

For some other cases, more can be said about the stability or instability of an
exact solution, as in \cite{becpre1}. However, those stability criteria are not
as straightforward to apply as the two given here, and we only make use
of the two criteria above. 

It follows from our derivation of \rf{eqn:noether} that the solution which has
a $\dn(m_i x_i,k_i)$ function in all directions is a deformation 
of the ground state of the
linear Schr\"odinger equation with periodic potential given by \rf{eqn:pot}.
Since \rf{eqn:nls} is nonlinear, the concept of a ground state as an
eigenfunction corresponding to the lowest eigenvalue as used in
quantum mechanics has to be abandoned. However, the ground state
can still be defined as the global minimizer (if one exists) 
of the Hamiltonian of the equation,
with the constraint $||\psi||^2=C$, for some constant $C$.
This Hamiltonian is 
\beq\la{eqn:hamiltonian}
{\cal H}(\psi)=\int_{\Omega} \left(\frac{1}{2} \nabla
\psi \cdot \nabla \psi^*+V(\vec{x})|\psi|^2+\frac{1}{2}\alpha |\psi|^4)\right)d\vec{x}. 
\eeq
\no It is bounded from below only for condensates with repulsive interactions.
Appendix A (due to Jared C. Bronski) demonstrates that in this case, the 
$\dn(m_i x_i,k_i)$ solution
is indeed the ground state of the nonlinear equation. Since the Hamiltonian
\rf{eqn:hamiltonian} is not bounded from below for condensates with attractive
interactions, no such ground state exists for this case. 


\section{Condensates with repulsive interaction}  

In this section, the numerical solutions of \rf{eqn:nls} in the repulsive
regime of the nonlinear Schr\"odinger equation are considered.  The initial
conditions are selected from the exact solutions given by \rf{eqn:sep} and
(\ref{eqn:density}-\ref{eqn:ci}).  Of particular interest are the trivial-phase
solutions given by (\ref{eqn:trivphasea}-c) and the nontrivial phase solutions
in the trigonometric limit \rf{eqn:trig}.  These solutions automatically
satisfy the phase quantization condition for our periodic domain
\cite{becpre2,becpre1} and correspond to a solution with a ramped phase
profile.  The numerical procedure used is a fourth-order Runge-Kutta method in
time and a filtered pseudospectral method in space \cite{numerics}.  For each
experiment, a small amount of white noise was added to the initial conditions
as a perturbation to expedite the onset of any instability.

\begin{figure}[tb]
\centerline{\psfig{figure=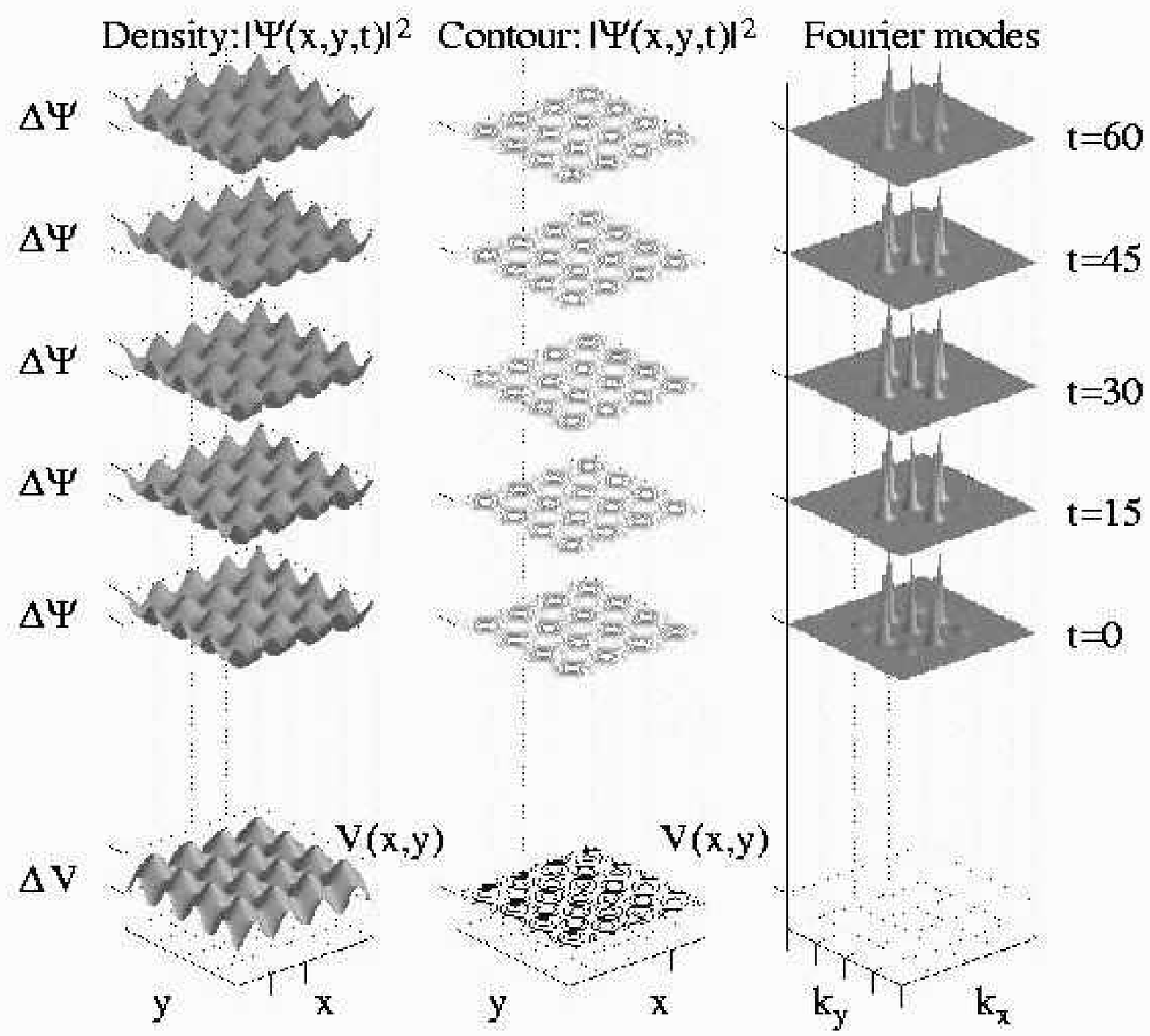,width=130mm,silent=}}
\begin{center}
\vspace*{-0.4in}
\caption{\label{fig:dnstck} The two-dimensional stable repulsive evolution of
 the  $\dn(m_i x_i,k_i)$ solution corresponding to \rf{eqn:trivphasec} over four
 periods, with $k_1=k_2=0.5$, $m_1=m_2=1$, and $A_1=A_2=-1$.} 
\end{center}
\end{figure}

\subsection{Numerical Simulations:  Square, Regular Lattice}

Our computational studies begin by considering the case of a
square, regular lattice.  Thus the spatial domains of all $x_i$ are
identical as are the elliptic moduli $k_i$, and the generated
lattice potential $V(\vec{x})$ and the solution $\psi(\vec{x},t)$ 
do not distinguish between the different directions $x_i$. 

%
\begin{figure}[tb]
\centerline{\psfig{figure=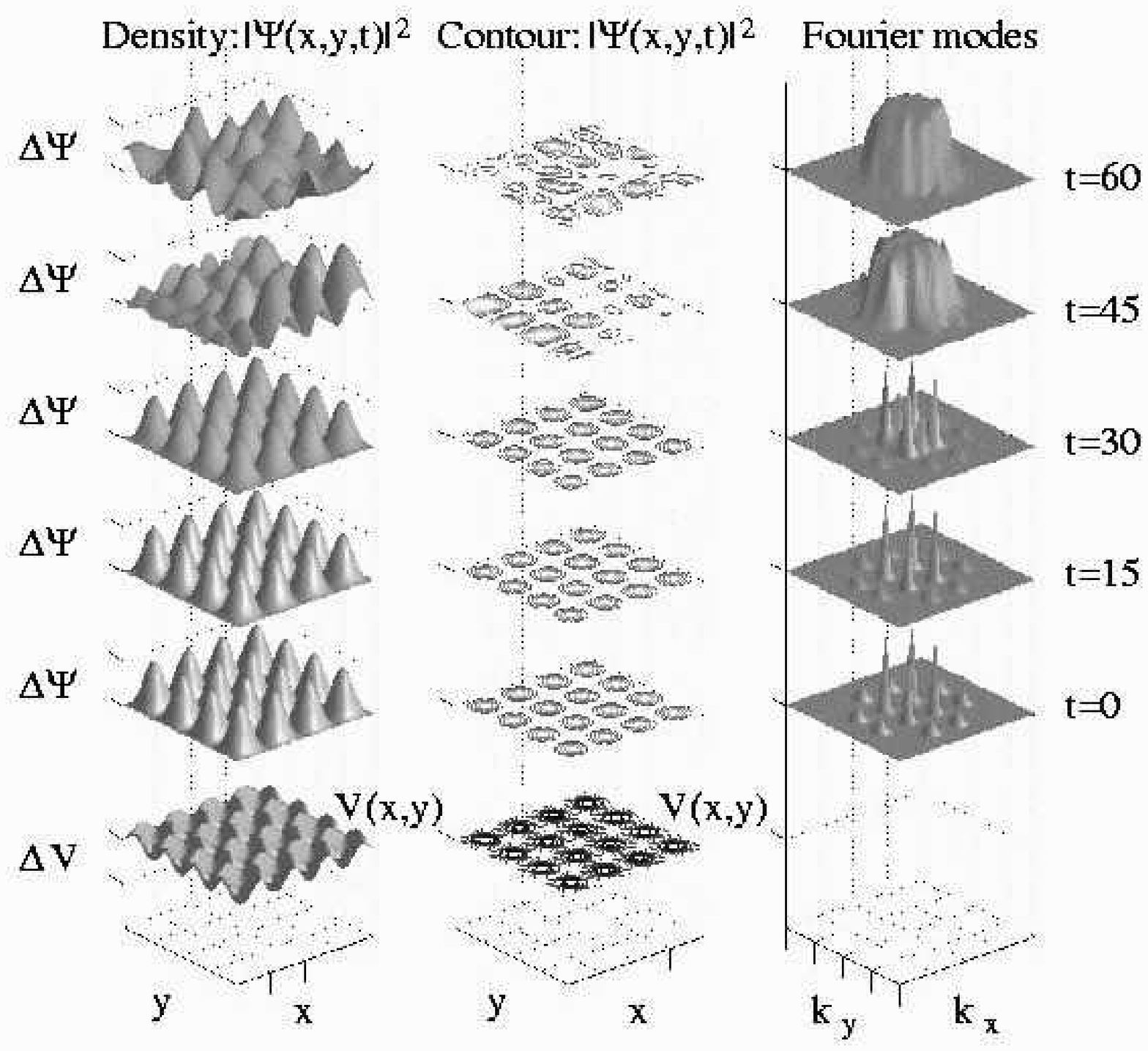,width=130mm,silent=}}
\begin{center}
\vspace*{-0.4in}
\caption{\label{fig:snstck} The two-dimensional unstable repulsive evolution
 of the  $\sn(m_i x_i,k_i)$ solution corresponding to \rf{eqn:trivphasea} over
 four periods, with $k_1=k_2=0.5$, $m_1=m_2=1$, and $A_1=A_2=1$.} 
\end{center}
\end{figure}
%

\subsubsection{Two-Dimensional Solutions}

The first two-dimensional solution we consider is the $\dn(m_i
x_i,k_i)$ solution \rf{eqn:trivphasec}.  As predicted by Theorem \ref{theo1}, 
this nodeless solution is linearly stable for
a condensate with repulsive interactions.  The evolution of four 
periods of this solution with $A_i=-1$, 
$m_i=1$ and $k=0.5$, $i=1,2$ for $t\in[0,60]$ is shown in
Fig. \ref{fig:dnstck}.  The three columns of this figure, and all other
figures of this type, represent in order the evolution of the density
$|\psi(x,y,t)|^2$, a contour plot of this same evolution, and the evolution of
the arctan of the
Fourier spectrum. The bottom picture in the first two columns shows the
potential $V(x,y)$. The arctan is applied to the
spectral evolution to limit the range of the power spectrum.  This
results in a more elucidating representation of the dynamics of the exact
solution, by removing the extreme contrasts in the range of the Fourier spectrum.
For the nodeless $\dn(m_i
x_i,k_i)$ solution, we note that over the time range considered, and
indeed for times $t\rightarrow \infty$, the solution does not change even when
strongly perturbed.  As argued previously \cite{becprl1,becpre1,pla}, the
stability of this solution is reminiscent of the stability of the
plane wave solution of the nonlinear Schr\"odinger equation with repulsive
nonlinearity
\cite{sulem}.

\begin{figure}[tb]
\centerline{\psfig{figure=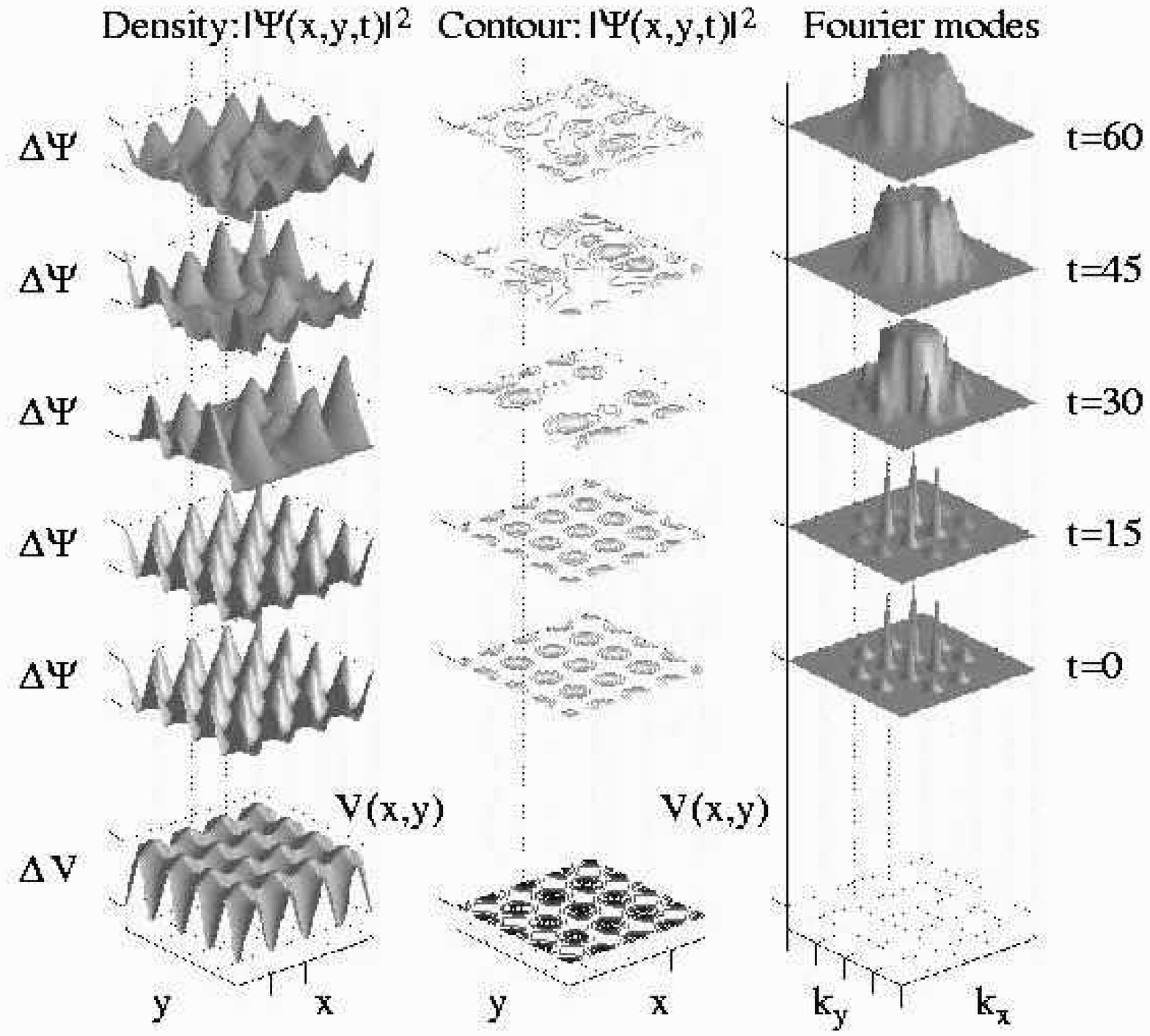,width=130mm,silent=}}
\begin{center}
\vspace*{-0.4in}
\caption{\label{fig:cnstck} The two-dimensional unstable repulsive evolution
 of the  $\cn(m_i x_i,k_i)$ solution corresponding to \rf{eqn:trivphaseb} over
 four periods, with $k_1=k_2=0.5$, $m_1=m_2=1$, and $A_1=A_2=-1$.} 
\end{center}
\end{figure}

In contrast to the stable $\dn(m_i x_i,k_i)$ solution,
the $\sn(m_i x_i,k_i)$ and $\cn(m_i x_i,k_i)$ solutions corresponding 
to \rf{eqn:trivphasea} and \rf{eqn:trivphaseb} are unstable.  
Both these solutions have nodes and violate the linear stability criterion of
Theorem \ref{theo1}.  The evolution for $t\in[0,60]$ of four periods 
of these solutions is shown in Figs. \ref{fig:snstck} and \ref{fig:cnstck}.  
The evolution of the density 
clearly shows the onset of a modulational instability which deforms
the exact solution.  The power spectrum shows that this modulational 
instability results in the activation of a large number of Fourier modes, 
destroying the possibility of stable evolution.

%
%

\begin{figure}[tb]
\centerline{\psfig{figure=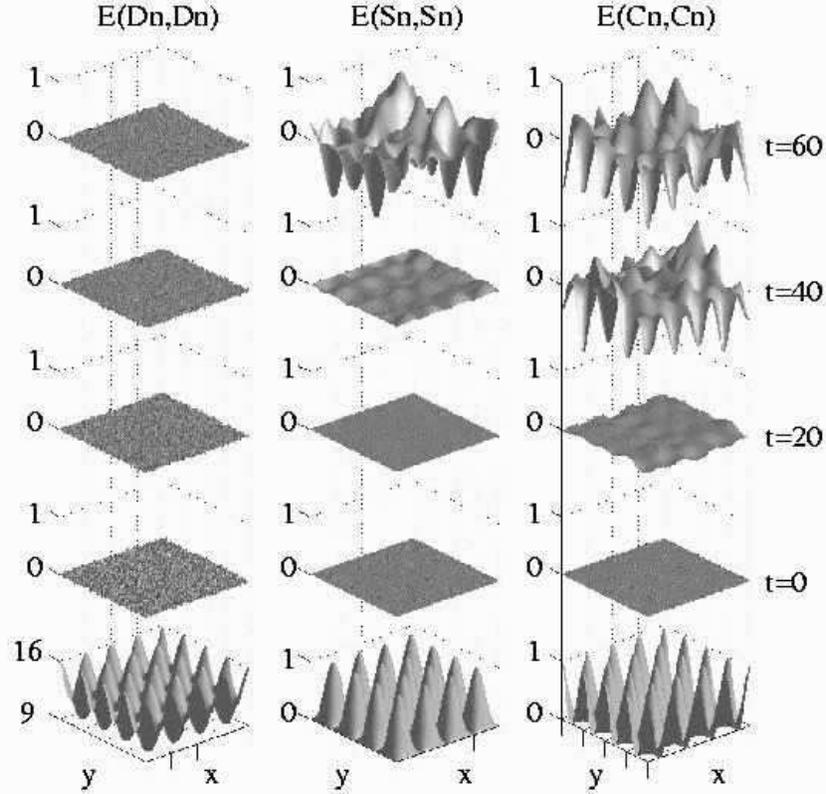,width=130mm,silent=}}
\begin{center}
\vspace*{-0.4in}
\caption{\label{fig:error} The two-dimensional repulsive evolution of the
 error, $E=|\psi(x,y,t)|^2-|\psi(x,y,0)|^2$, for the 
 $\sn(m_i x_i,k_i)$, $\cn(m_i x_i,k_i)$ and $\dn(m_i x_i,k_i)$ 
 solutions of \rf{eqn:trivphasea}-\rf{eqn:trivphasec}.  The solutions
 considered correspond to Figs. \ref{fig:dnstck}-\ref{fig:cnstck}.
 The bottom row shows the respective exact solutions.}
\end{center}
\end{figure}

To illustrate the computational stability results, we calculate the difference
$E$ between the density of the exact solution and the density of the numerical
(perturbed) solution for the $\sn(m_i x_i,k_i)$, $\cn(m_i x_i,k_i)$ and
$\dn(m_i x_i,k_i)$  solutions of \rf{eqn:trivphasea}-\rf{eqn:trivphasec}.
Thus,  $E=|\psi(x,y,t)|^2-|\psi(x,y,0)|^2$ . If the numerical evolution was
exact and the initial condition was unperturbed by white noise, this difference
would be identically zero, since the solutions we consider are stationary. Thus
any growth in this difference is due to an instability mechanism which causes
any errors to grow.  
In Fig. \ref{fig:error}, the three columns given by $E(Dn,Dn)$, $E(Sn,Sn)$, and
$E(Cn,Cn)$ represent the errors in the exact solutions given by the $\dn(m_i
x_i,k_i)$, $\sn(m_i x_i,k_i)$ and $\cn(m_i x_i,k_i)$  solutions of
\rf{eqn:trivphasea}-\rf{eqn:trivphasec} respectively. For the $\dn(m_i
x_i,k_i)$ solution, the error $E(Dn,Dn)$ remains at the level of the initial
noise.   In contrast, the errors for the $\sn(m_i x_i,k_i)$ and $\cn(m_i
x_i,k_i)$ solutions, $E(Sn,Sn)$ and $E(Cn,Cn)$, start to grow at the onset of
instability near $t\approx 40$ and $t\approx 20$  respectively.  

%
%

\begin{figure}[tb]
\hspace*{-1cm}\centerline{\psfig{figure=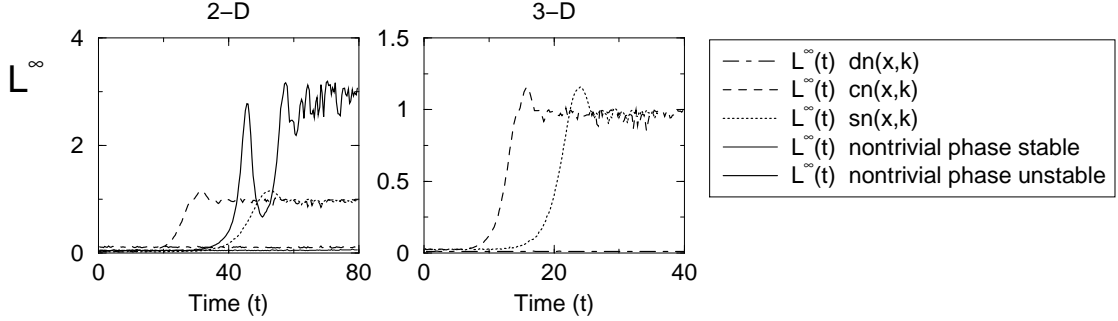,width=120mm,silent=}}
\begin{center}
\vspace*{-0.3in}
\caption{\label{fig:Linfdef} Evolution of the $L^\infty$ norm of the error 
   in two- 
   and three-dimensions for several trivial-phase solutions of the repulsive
   equation.} 
\end{center}
\end{figure}

\sloppypar 
The unstable behavior is further illustrated by the evolution of the $L^\infty$
norm of $E$:  ${\rm max}_{\{x,y\}}\left| |\psi(x,y,t)|^2-|\psi(x,y,0)|^2 \right|$.  In Fig.
\ref{fig:Linfdef}, this $L^\infty$ norm dynamics is given for both the
two-dimensional and three-dimensional solutions of the equation with repulsive
nonlinearity 
considered numerically in this paper. The $L^\infty$ norm of the unstable
solutions grows at the onset of the instability and saturates at  a finite
value.  This reflects the nature of the repulsive instability, {\em i.e.}, no
large gradients or sharp peaks are allowed to develop in the solution. For the
stable solutions, this $L^\infty$ norm remains at the initial noise level. 

%
%

\subsubsection{Three-Dimensional Solutions}

\begin{figure}[tb]
\vspace*{-0.2in}
\centerline{\psfig{figure=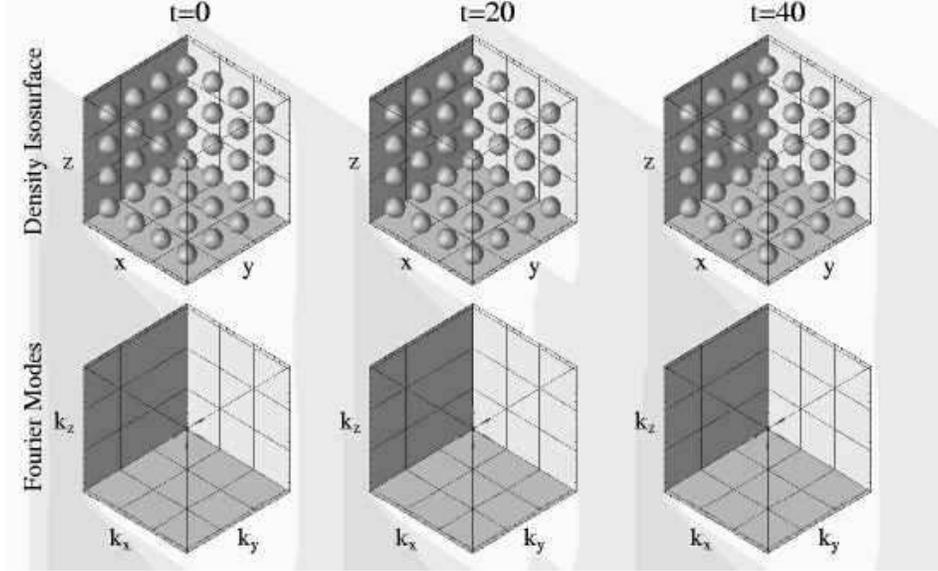,width=125mm,silent=}}
\begin{center}
\vspace*{-0.1in}
\caption{\label{fig:dndefocus}
 The three-dimensional stable repulsive evolution of
 the  $\dn(m_i x_i,k_i)$ solution corresponding to \rf{eqn:trivphasec} 
 over eight
 periods (two in each direction), 
 with $k_1=k_2=k_3=0.5$, $m_1=m_2=m_3=1$, and $A_1=A_2=A_3=-0.5$.
 The Fourier spectrum is composed of six symmetric peaks around the origin 
 which are obscured by the grid lines.}
\end{center}
\end{figure}

As in the two-dimensional case, we first consider the
trivial phase $\dn(m_i x_i,k_i)$ solution \rf{eqn:trivphasec}. As shown in
Theorem \ref{theo1},
this nodeless solution is linearly stable.
The evolution of eight periods of
this solution with $A_i=-0.5$, $m_i=1$ and $k=0.5$, $i=1,2,3$ for $t\in[0,40]$
is shown in Fig. \ref{fig:dndefocus}.  The two rows of this figure, and all other
figures of this type, represent the evolution of a selected density 
iso-surface
$|\psi(x,y,z,t)|^2=$ constant, and the evolution of an iso-surface of 
the arctan of the Fourier spectrum.
Unless otherwise stated, the density iso-surface shown is at $30\%$
of the value from minimum to maximum while the iso-surface of the Fourier
spectrum is at a value of one.  Note that the arctan of the spectrum
limits the value of the power density to a maximum of $\pi/2$.  
Over the time range considered, and
indeed for times $t\rightarrow \infty$,
the nodeless
$\dn(m_i x_i,k_i)$ solution is unaffected by the
perturbations. This even holds for  
perturbations that are strong compared to the amplitude of the solution. 

\begin{figure}[tb]
\vspace*{-0.2in}
\centerline{\psfig{figure=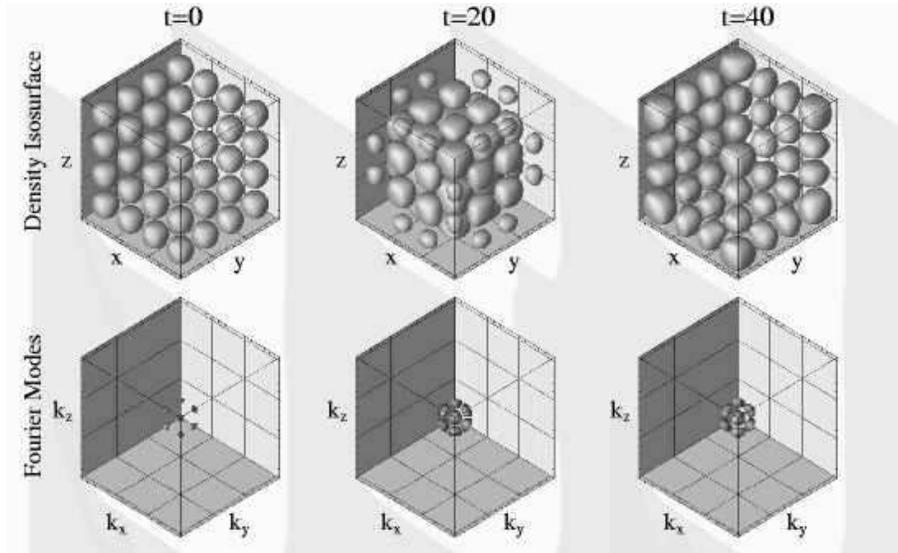,width=120mm,silent=}}
\begin{center}
\vspace*{-0.1in}
\caption{\label{fig:sndefocus} The three-dimensional unstable repulsive evolution
 of the  $\sn(m_i x_i,k_i)$ solution corresponding to \rf{eqn:trivphasea} over
 eight periods (two in each direction), with $k_1=k_2=k_3=0.5$, 
 $m_1=m_2=m_3=1$, and $A_1=A_2=A_3=1$.  At $t=40$ the further loss of
 structure, which is exemplified by the Fourier spectrum, 
 is hidden by the shape of the iso-surfaces at 
 the boundary of the box.} 
\end{center}
\end{figure}

\begin{figure}[tb]
\centerline{\psfig{figure=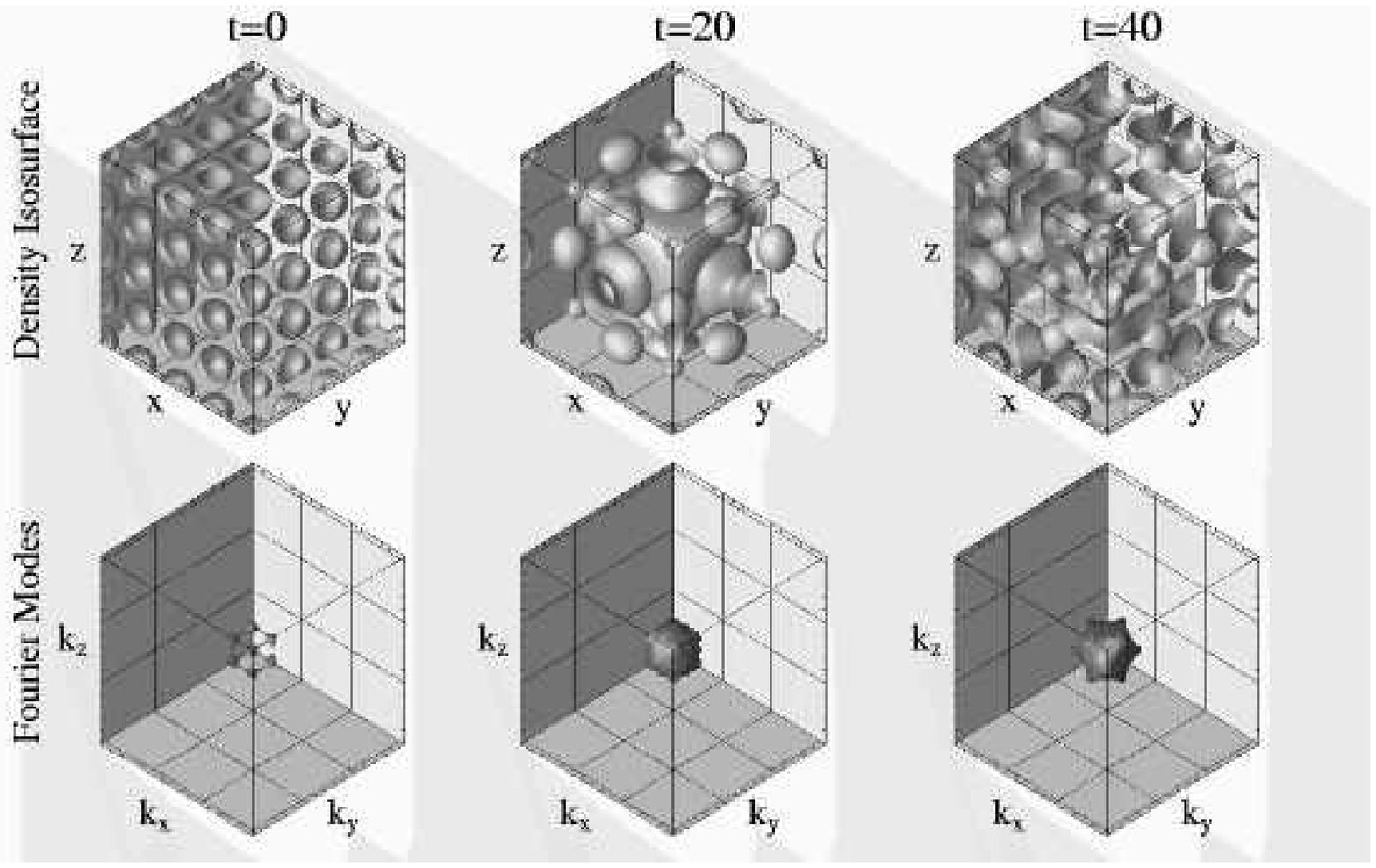,width=120mm,silent=}}
\begin{center}
\caption{\label{fig:cndefocus} The three-dimensional unstable repulsive evolution
 of the  $\cn(m_i x_i,k_i)$ solution corresponding to \rf{eqn:trivphaseb} over
 eight periods (two in each direction), with $k_1=k_2=k_3=0.5$, $m_1=m_2=m_3=1$, and $A_1=A_2=A_3=-1$.}
\end{center}
\end{figure}

Just as in the two-dimensional case, the $\sn(m_i x_i,k_i)$ and $\cn(m_i
x_i,k_i)$ solutions corresponding to (\ref{eqn:trivphasea}-b)
are unstable.  Both these solutions have nodes and violate
the stability criterion of Theorem \ref{theo1}.  The
evolution for $t\in[0,40]$ of eight periods of these solutions is shown in Figs.
\ref{fig:sndefocus} and \ref{fig:cndefocus}.  The evolution of the density
clearly shows the onset of a modulational instability which causes break-up of 
the exact
solution.  The Fourier spectrum of the density 
shows that this modulational instability results
in the activation of a large number of Fourier modes, destroying the
possibility of stable evolution.  Fig. \ref{fig:Linfdef} illustrates
this unstable behavior by showing the evolution of the $L^\infty$ norm of the
error:
${\rm max}_{\{x,y,z\}}\left| |\psi(x,y,z,t)|^2-|\psi(x,y,z,0)|^2 \right|$.  As with
the two-dimensional results, the $L^\infty$ norm of the error of the 
unstable solutions
grows at the onset of the instability and saturates at a finite value.  For
the stable solutions, this $L^\infty$ norm remains at the initial noise level.

%
%

%
%

\clearpage

%
%

\subsection{Numerical Simulations:  Rectangular, Irregular Lattice}

\begin{figure}[tb]
\centerline{\psfig{figure=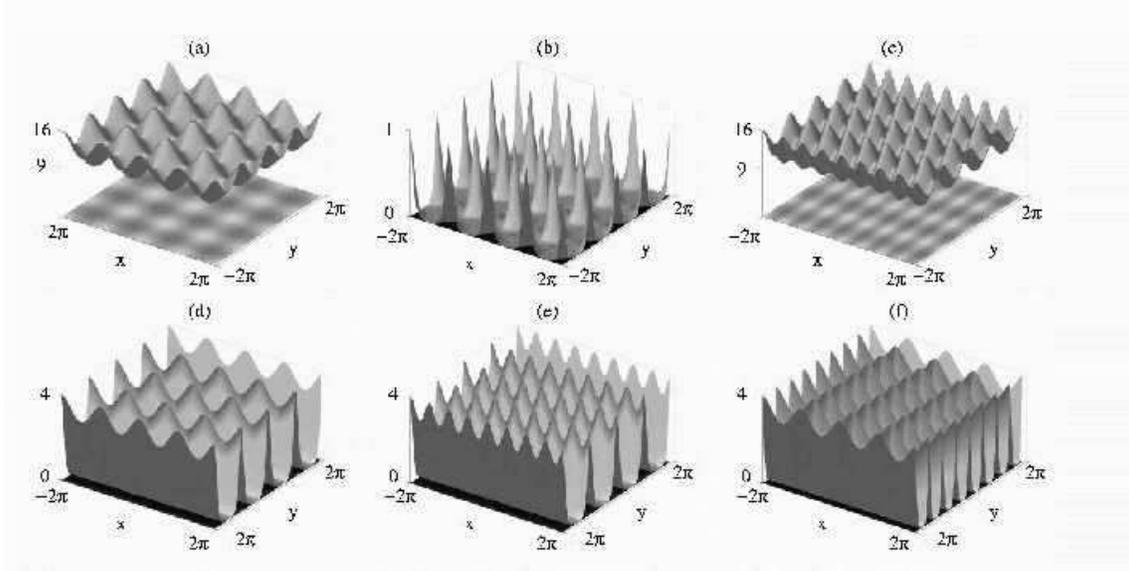,width=150mm,silent=}}
\begin{center}
\caption{\label{fig:dnpics} 
Stable two-dimensional solutions with irregular
lattice potentials and $m_i$ values chosen so that the solutions
are $4\pi$-periodic. (a) Two periods in each direction, with $k_1=0.1$ and $k_2=0.1$;
(b) two periods in each direction, with $k_1=0.9$ and 
$k_2=0.9$; 
(c) four periods in $x$ and two periods in $y$, with
  $k_1=0.1$ and $k_2=0.1$ respectively; (d) two periods in each
direction, with $k_1=0.1$ and $k_2=0.9$;
(e) four periods in $x$ and two periods in $y$, with
 $k_1=0.1$ and $k_2=0.9$ respectively; 
(f) two periods in $x$ and four periods in $y$, with
 $k_1=0.1$ and $k_2=0.9$ respectively.  The spatial domains are normalized 
  to $x,y\in[-2\pi,2\pi]$.} 
\end{center}
\end{figure}

In addition to regular lattices, we can consider more complicated
solutions by modifying the elliptic modulus $k_i$ and periodicity
parameter $m_i$.  We restrict ourselves to stable two-dimensional dynamics
given by the $\dn(m_i x_i,k_i)$ solution of \rf{eqn:trivphasec} since
they are easy to illustrate and stable solutions are the most relevant for
experiments.  The three-dimensional behavior
follows in a straightforward manner 
from a generalization of these two-dimensional visualizations.  
We only show the stable solutions at the initial time since they
are unchanged as $t\rightarrow\infty$.  Figure \ref{fig:dnpics}
displays the various behaviors as the elliptic modulus is
varied from $k=0.1$ to $k=0.999$ for various periods and values of
$m_i$.  The different stable solutions vary from periodic
lattice solutions in Fig. \ref{fig:dnpics}a and \ref{fig:dnpics}c,
to the well-separated and localized spikes of \ref{fig:dnpics}b, to
well-separated regions of oscillation and localization \ref{fig:dnpics}d-f.
Thus stability is independent of the lattice structure, as long as the solution
is off-set from the zero level.

%
%

\subsection{Numerical Simulations:  Nontrivial Phase}

\begin{figure}[tb]
\centerline{\psfig{figure=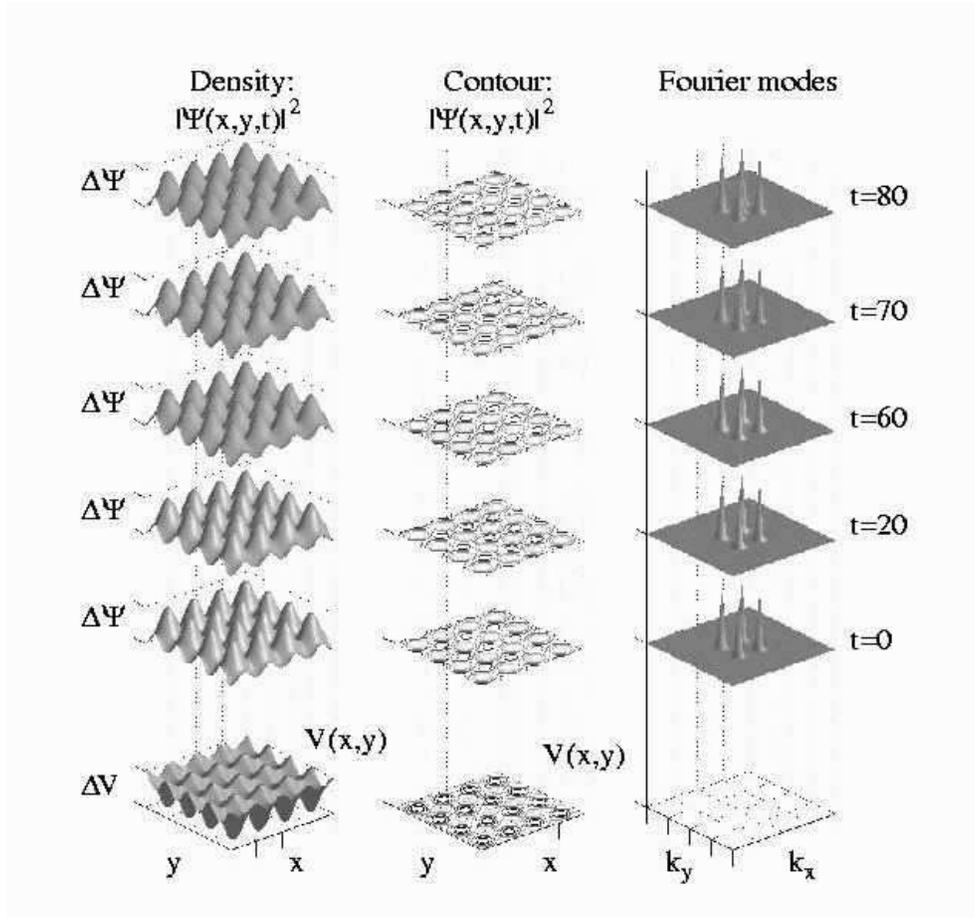,width=130mm,silent=}}
\begin{center}
\vspace*{-0.4in}
\caption{\label{fig:s2Dntpdefocus}
 The two-dimensional stable defocusing evolution of the nontrivial phase
 solution \rf{eqn:trig} with four
 periods, and $k_1=k_2=0.0$, $m_1=m_2=1$, $A_1=A_2=1$, $B_1=1$, and $B_2=0.7$.}
\end{center}
\end{figure}

\begin{figure}[tb]
\centerline{\psfig{figure=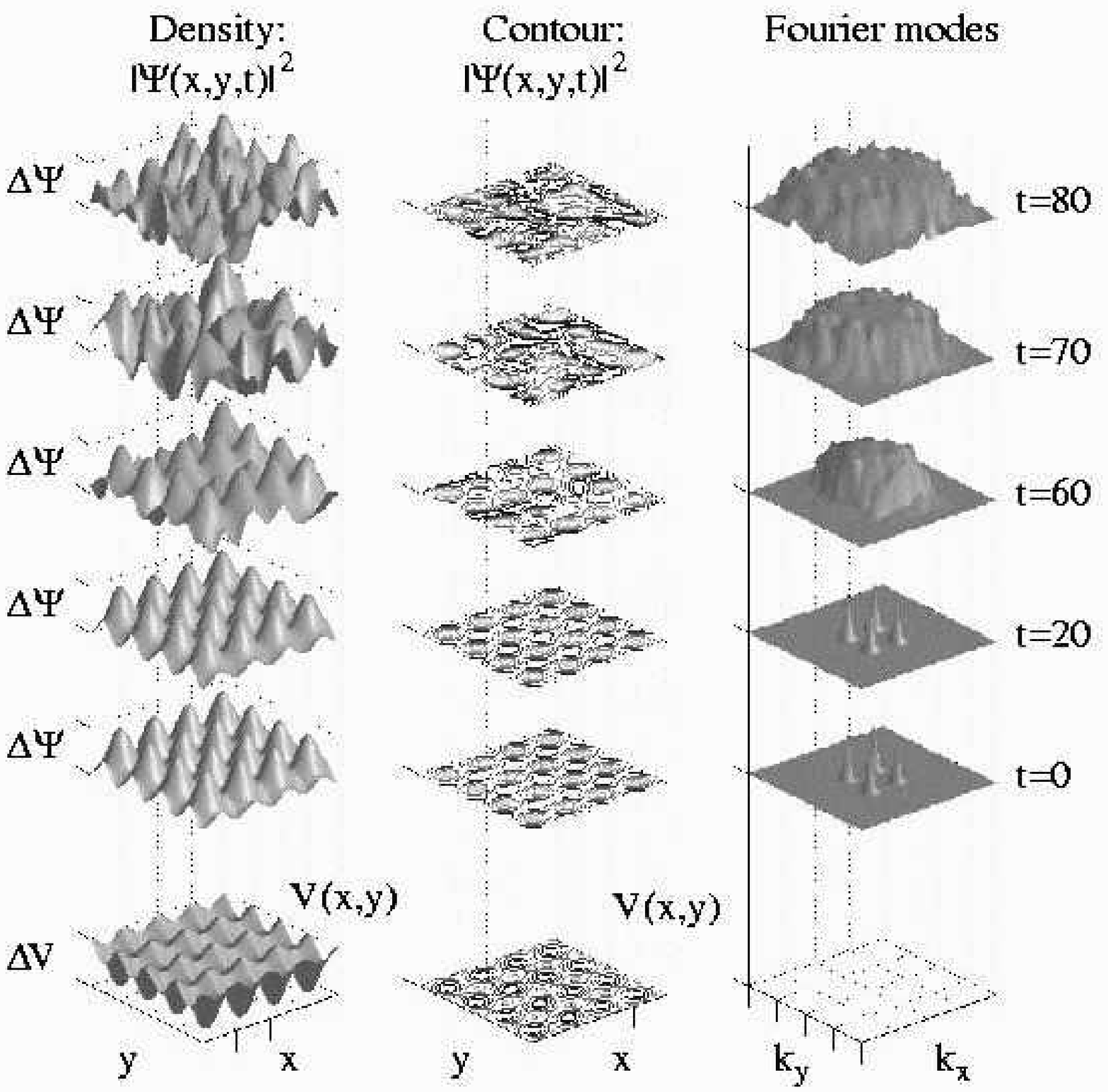,width=130mm,silent=}}
\begin{center}
\vspace*{-0.4in}
\caption{\label{fig:u2Dntpdefocus}
 The two-dimensional unstable repulsive evolution of the nontrivial phase
 solution \rf{eqn:trig} with four
 periods, with $k_1=k_2=0.0$, $m_1=m_2=1$, $A_1=A_2=1$, $B_1=1$, and $B_2=0.6$.}
\end{center}
\end{figure}

Analytical results for the stability of the nontrivial phase solutions are
difficult to obtain even in one dimension.  Thus we rely on numerical
investigations of the stability of these solutions.  To avoid the
complications which arise from phase quantization, consider the solutions
\rf{eqn:trig} that are trigonometric in all directions and for which phase
quantization is automatically satisfied.  For the NLS equation with repulsive
nonlinearity
\rf{eqn:nls} in two- and
three-dimensions, these solutions are stable or unstable depending on the
offset parameters $B_1$ and $B_2$ (two-dimensions) or $B_1, B_2$ and $B_3$
(three-dimensions).  As shown in Fig.~\ref{fig:s2Dntpdefocus} the off-set
solution in two-dimensions, which is qualitatively like the stable
$\dn(m_ix_i, k_i)$ solution, is stable with $B_1=1$ and $B_2=0.7$ whereas in
Fig.~\ref{fig:u2Dntpdefocus} the unstable nontrivial phase solution which is
below the offset threshold is illustrated with $B_1=1$ and $B_2=0.6$.  These
nontrivial phase solutions illustrate the necessity of offset for stability.  
In particular, for the values considered, an offset threshold is achieved at
$B_2\approx 0.65$.  

%
%

%
%

\begin{figure}[tb]
\vspace*{-.2in}
\centerline{\psfig{figure=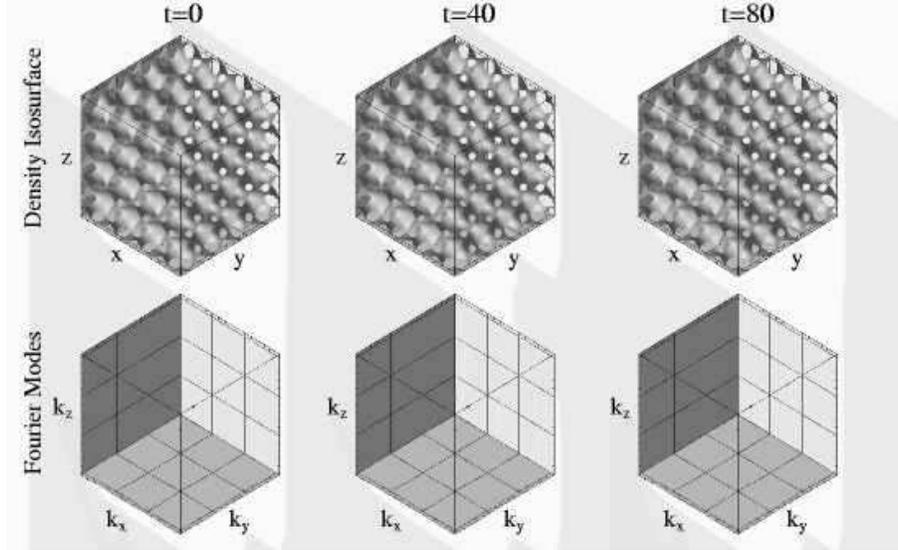,width=120mm,silent=}}
\begin{center}
\caption{\label{fig:sntpdefocus}
 The three-dimensional stable repulsive evolution of the nontrivial phase
 solution \rf{eqn:trig} with eight
 periods (two in each direction), 
 and $k_1=k_2=k_3=0.0$, $m_1=m_2=m_3=1$, $A_1=A_2=A_3=1$, and
 $B_1=B_2=B_3=1$.  The Fourier spectrum is composed of six symmetric 
 peaks around the origin 
 which are obscured by the grid lines.}
\end{center}
\end{figure}

\begin{figure}[tb]
\centerline{\psfig{figure=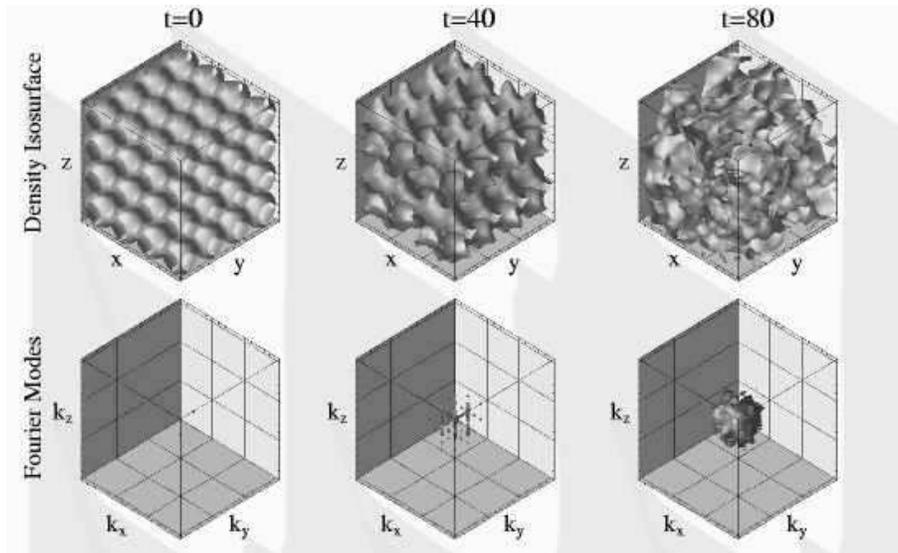,width=120mm,silent=}}
\begin{center}
\caption{\label{fig:untpdefocus}
 The three-dimensional unstable repulsive evolution of the nontrivial phase
 solution \rf{eqn:trig} with eight
 periods 
 (two in each direction), 
 with $k_1=k_2=k_3=0.0$, $m_1=m_2=m_3=1$, $A_1=A_2=A_3=1$, and
 $B_1=B_2=B_3=0.5$.}
\end{center}
\end{figure}

Similar results for repulsive nontrivial phase solutions hold in three
dimensions: the offset parameter $B_i$ is the crucial
parameter which determines the stability of a given nontrivial
phase solution.  Figure~\ref{fig:sntpdefocus} shows the stability of an off-set
solution in three-dimensions with $B_1=B_2=B_3=1$ whereas in
Fig.~\ref{fig:untpdefocus} an unstable nontrivial phase solution, which is
below the instability threshold, is illustrated with $B_1=B_2=B_3=0.5$. 

\begin{figure}[tb]
\vspace*{-.2in}
\centerline{\psfig{figure=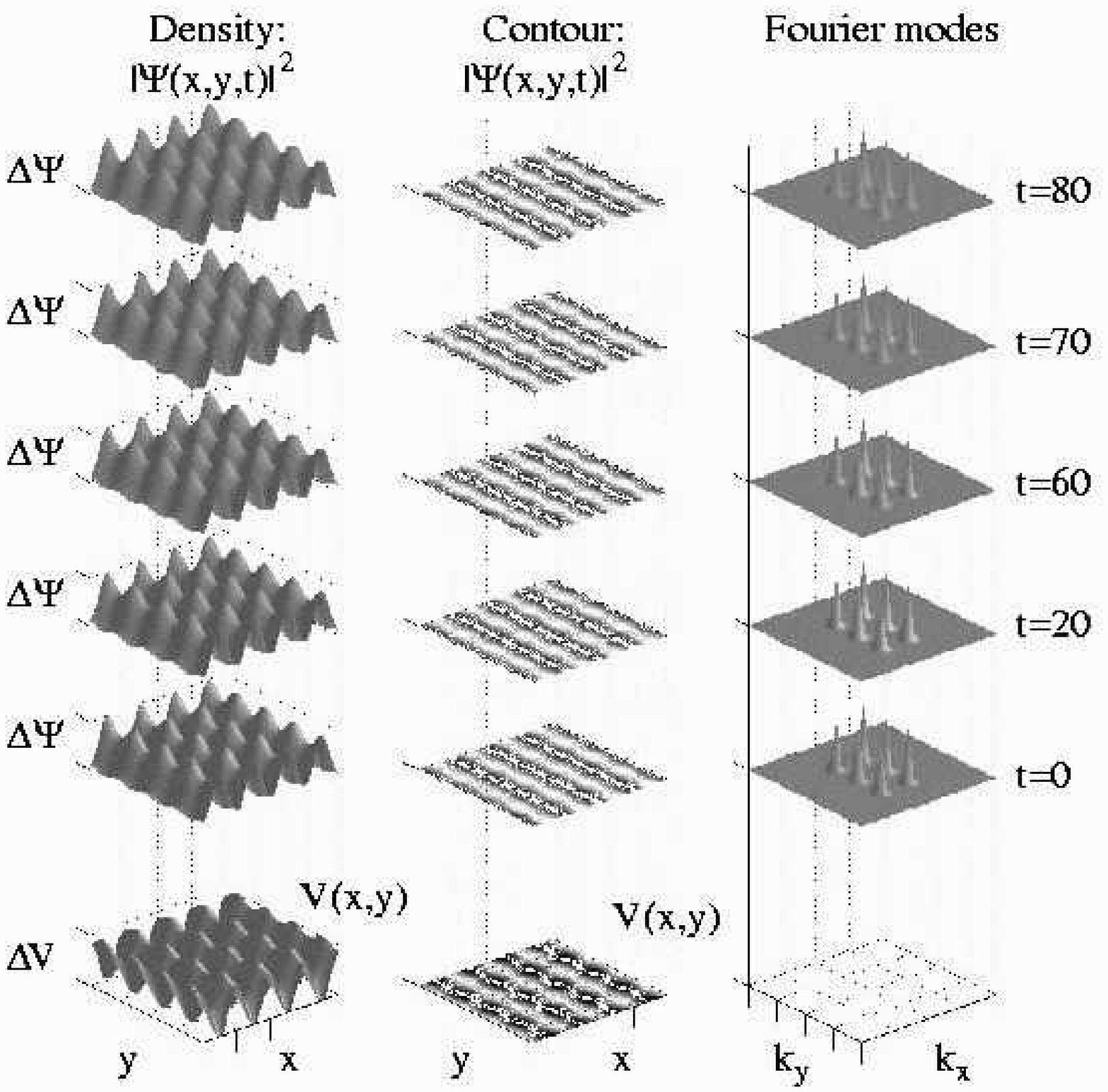,width=130mm,silent=}}
\begin{center}
\vspace*{-.4in}
\caption{\label{fig:sdn-ntpdefocus}
 The two-dimensional stable repulsive evolution of a mixed solution
 which has nontrivial phase profile \rf{eqn:trig} in the $x$ direction 
 and the $\dn(m_i x_i,k_i)$ profile of \rf{eqn:trivphasec} in the $y$
 direction.   The figure shows two periods in each direction
 with $k_1=0$, $k_2=0.5$, $m_1=m_2=1$, $A_1=1$, $A_2=-0.5$, and $B_1=0.4$.}
\end{center}
\end{figure}

\begin{figure}[tb]
\vspace*{-.2in}
\centerline{\psfig{figure=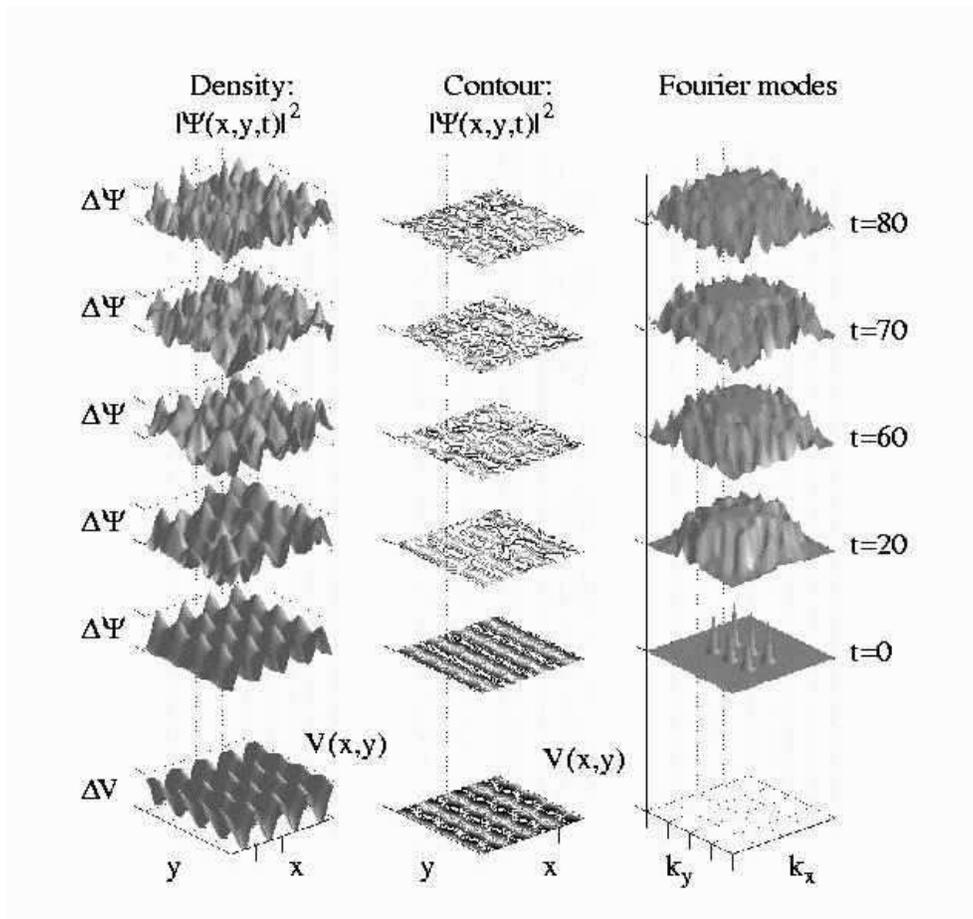,width=130mm,silent=}}
\begin{center}
\vspace*{-.4in}
\caption{\label{fig:udn-ntpdefocus}
 The two-dimensional unstable repulsive evolution of a mixed solution
 which has a nontrivial phase profile \rf{eqn:trig} in the $x$ direction 
 and the $\dn(m_i x_i,k_i)$ profile of \rf{eqn:trivphasec} in the $y$
 direction.   The figure shows two periods
 with $k_1=0$, $k_2=0.5$, $m_1=m_2=1$, $A_1=1$, $A_2=-0.5$, and $B_1=0.3$.}
\end{center}
\end{figure}

Mixed solutions can also be considered:  these are solutions with nontrivial
phase \rf{eqn:trig} in one or more directions and trivial phase
\rf{eqn:trivphasea} in the remaining directions.  As before,
offset determines the stability of these solutions.  To
illustrate this, we consider the two-dimensional repulsive evolution 
with a trivial phase $\dn(m_i x_i,k_i)$ solution of 
\rf{eqn:trivphasec} in the $y$ direction and a nontrivial phase solution
\rf{eqn:trig} in the $x$ direction.  The $\dn(m_i x_i,k_i)$ solution of 
\rf{eqn:trivphasec} was found to be stable whereas the nontrivial phase solution
\rf{eqn:trig} was stable provided a sufficient offset was present.
Following the previous paragraphs, we shown in Fig.~\ref{fig:sdn-ntpdefocus} 
the mixed solution with offset parameter $B_1=0.4$ for the nontrivial phase
dimension.  The resulting evolution is stable under perturbation as 
$t\rightarrow\infty$.  Alternatively, when the offset parameter is
decreased to $B_1=0.3$, the nontrivial phase solution is unstable, as 
observed in Fig.~\ref{fig:udn-ntpdefocus}.

\section{Condensates with attractive interaction}

In this section, the numerical solutions in the attractive regime of the
nonlinear Schr\"odinger equation \rf{eqn:nls} are considered.  The initial
conditions are only selected from the exact trivial-phase solutions given by
(\ref{eqn:trivphasea}-c).  These trivial phase solutions, given by the
$\sn(m_i x_i,k_i)$, $\cn(m_i x_i,k_i)$ and $\dn(m_i x_i,k_i)$ ($i=1, 2,$ or
$3$) of (\ref{eqn:trivphasea}-c), characterize the basic solution types:
peak-on-peak with nodes, peak-on-peak without nodes, peak-on-trough with
nodes, and peak-on-trough without nodes.

It is well known \cite{sulem} that in two dimensions and higher, solutions of
the attractive nonlinear Schr\"odinger equation undergo a process of collapse
and blow-up.  This process is also possible in the presence of a periodic potential.
It is briefly discussed in Appendix B.  In addition, plane wave solutions are
known to be modulationally unstable \cite{sulem}.  Thus we do not expect
stable two- and three-dimensional solutions.  However, it may be possible to
obtain a solution for which the onset of instability occurs on a longer
time scale than the experimental lifetime of a condensate.  This possibility is
examined numerically.

\subsection{Numerical Simulations:  Square, Regular Lattice}

Our numerical considerations begin with the case of a square, regular lattice.
Thus the spatial domains of all $x_i$ are identical as are the elliptic moduli
$k_i$.  Thus, the generated lattice potential $V(\vec{x})$ and the solution
$\psi(\vec{x},t)$ do not distinguish between the different directions $x_i$.

%
%

%
%

\clearpage

%
%

%
%

\subsubsection{Two-Dimensional Solutions}

\begin{figure}[tb]
\vspace*{-0.2in}
\centerline{\psfig{figure=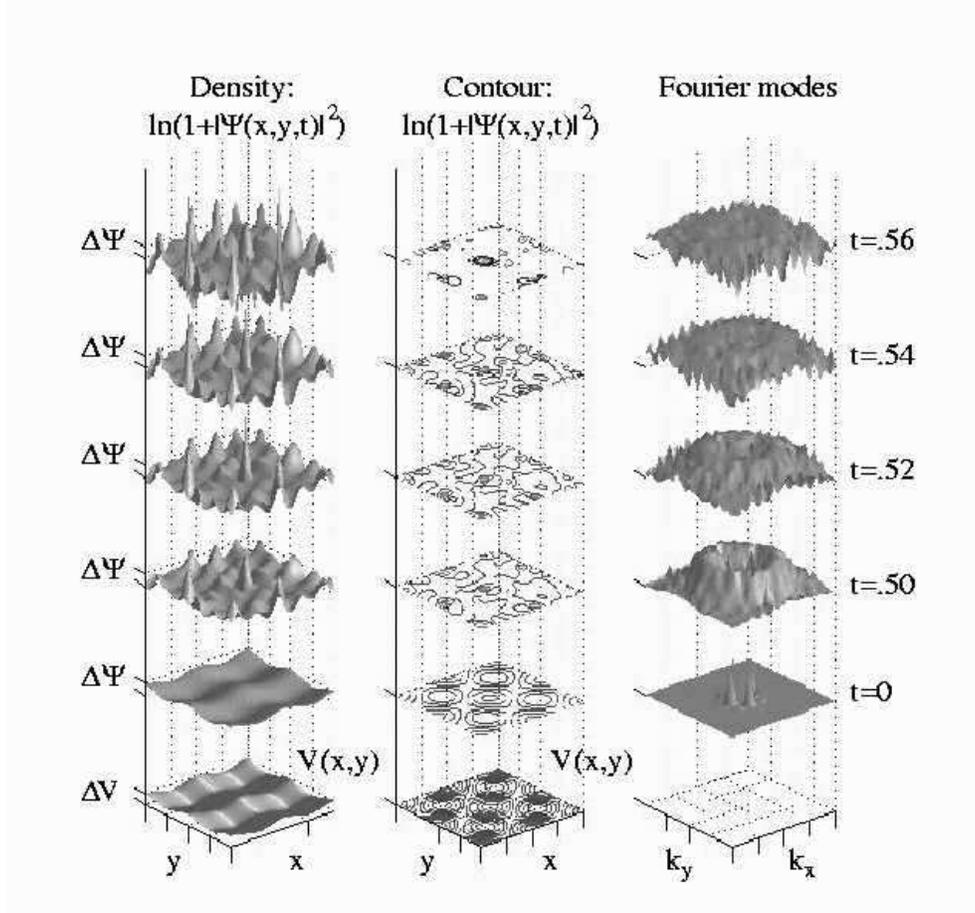,width=130mm,silent=}}
\begin{center}
\vspace*{-0.4in}
\caption{\label{fig:dn-dn-stack}
 The two-dimensional unstable attractive evolution of
 the double-amplitude $\dn(m_i x_i,k_i)$ initial condition corresponding to \rf{eqn:trivphasec} over four
 periods, with $k_1=k_2=0.5$, $m_1=m_2=1$, and $A_1=A_2=-1$.}
\end{center}
\end{figure}

The first two-dimensional solution considered is the $\dn(m_i x_i,k_i)$
solution \rf{eqn:trivphasec}.  For the attractive nonlinear Schr\"odinger
equation, the $\dn(m_i x_i,k_i)$ solution is proved to be unstable in 
Theorem~\ref{theo2}.  This is a manifestation of the modulational instability
of plane waves \cite{sulem}.  Although the modulational instability is present
in the dynamics, the behavior is dominated by the collapse and blow-up
phenomena which are characteristic of the attractive regime.  This behavior is
observed in Fig. \ref{fig:dn-dn-stack}.  In this case, 
$\ln (1+|\psi(x,y,t)|^2)$, and its contours are plotted to better illustrate the
collapse and blow-up.  The evolution of four periods of this solution with
$A_i=-1$, $m_i=1$ and $k=0.5$, $i=1,2$ for $t\in[0,0.56]$ shows how quickly
this phenomenon occurs.  

In order to effectively capture the onset of collapse, we have aided the
instability in some simulations by considering initial conditions which have
twice the amplitude of an exact solution.  Our motivation in expediting the
instability is to limit computational cost: since the exact solutions are
stationary, sufficient perturbation is required to achieve collapse. Building
up such perturbations from our regular white-noise perturbations requires long
time scales relative to the time scale on which collapse occurs (on the
order of 1000 times longer for our numerical experiments). In order to resolve
the collapse, a small time step is required, which is unnecessary for the time
leading up to collapse. An alternative approach to avoid this problem is to
use an adaptive time-stepping algorithm. The main effect of our approach 
is to alter the time of collapse and blow-up.  
Other qualitative features of the solution
are unaffected.

Note that the double-amplitude $\dn(m_i x_i,k_i)$ initial condition, which is
nodeless and has a larger $L^2$ norm than either the $\sn(m_i x_i,k_i)$ or
$\cn(m_i x_i,k_i)$ solution, collapses into a few distinct peaks. Additionally,
the spreading of the Fourier mode spectrum reflects the nature of the
localization inherent to collapse. 

%
%

\begin{figure}[tb]
\vspace*{-0.2in}
\centerline{\psfig{figure=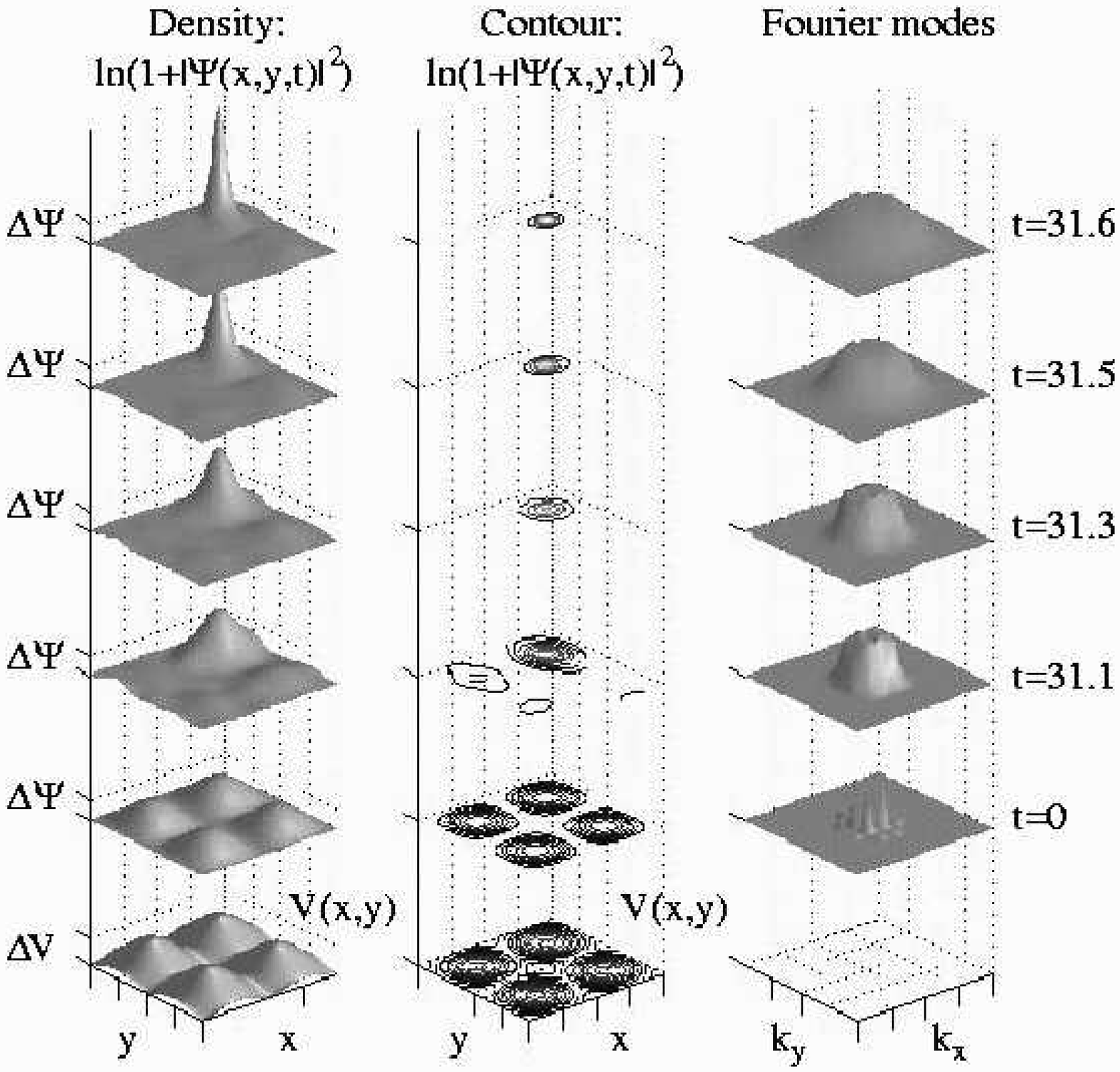,width=130mm,silent=}}
\begin{center}
\vspace*{-0.4in}
\caption{\label{fig:sn-sn-stack}
 The two-dimensional unstable attractive evolution of
 the  $\sn(m_i x_i,k_i)$ solution corresponding to \rf{eqn:trivphasec} over four
 periods, with $k_1=k_2=0.5$, $m_1=m_2=1$, and $A_1=A_2=1$.}
\end{center}
\end{figure}

The exact peak-on-peak $\sn(m_i x_i,k_i)$ solution is also
unstable for attractive condensates as shown in Fig. \ref{fig:sn-sn-stack}.  As
with the $\dn(m_i x_i,k_i)$ initial condition, the solution eventually blows up near
$t\approx 31$.  In this case, the total $L^2$ norm is much smaller than that
of the $\dn(m_i x_i,k_i)$ solution so that only a single collapse peak is
observed.  In contrast to Fig. \ref{fig:dn-dn-stack}, the initial conditions
are the exact solutions with a small amount of white noise added.

%
%

\begin{figure}[tb]
\vspace*{-0.2in}
\centerline{\psfig{figure=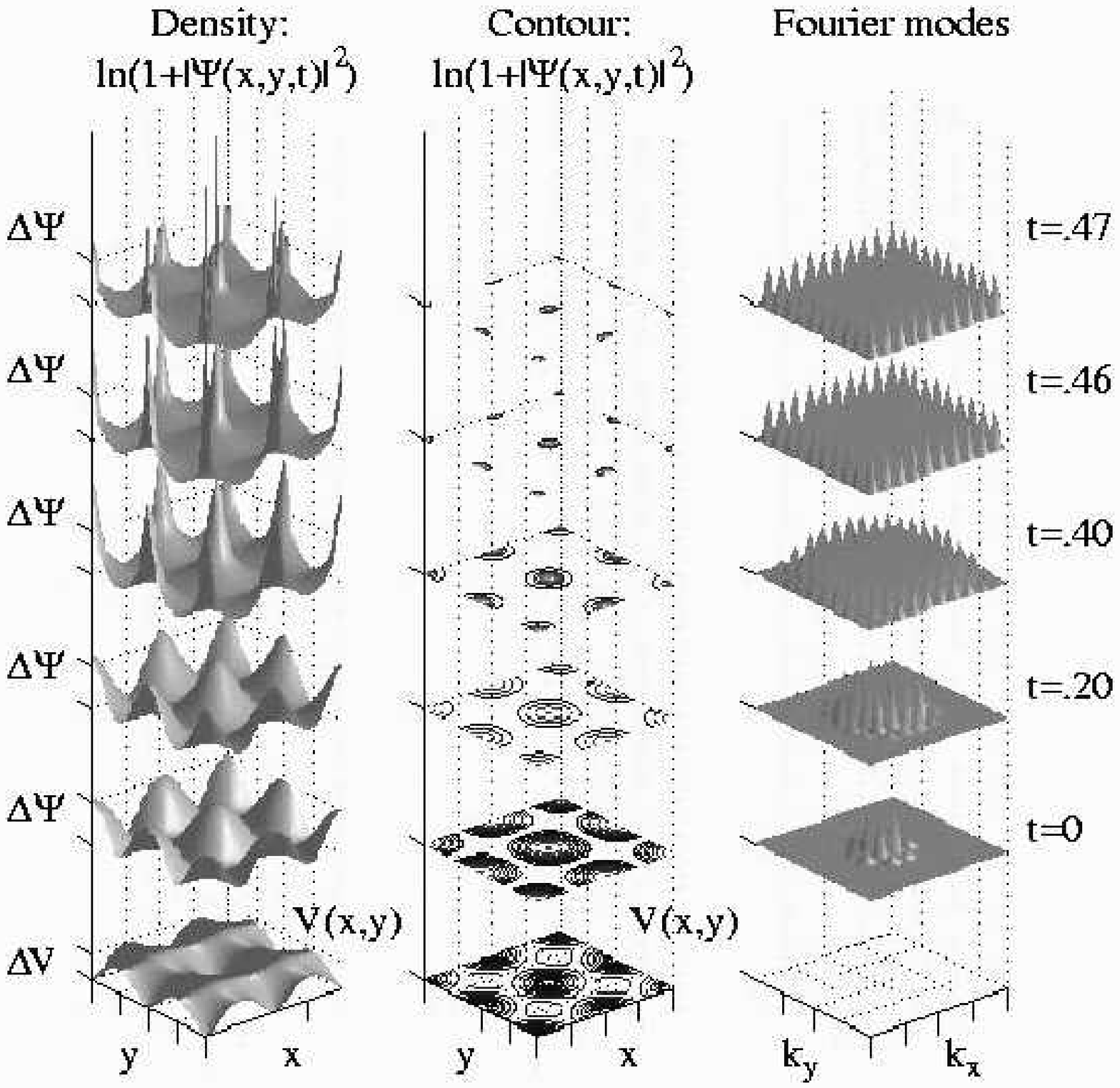,width=130mm,silent=}}
\begin{center}
\vspace*{-0.4in}
\caption{\label{fig:cn-cn-stack}
 The two-dimensional unstable attractive evolution of
 the  double-amplitude $\cn(m_i x_i,k_i)$ initial condition 
 corresponding to \rf{eqn:trivphasec} with four
 periods, and $k_1=k_2=0.5$, $m_1=m_2=1$, and $A_1=A_2=-1$.}
\end{center}
\end{figure}

\begin{figure}[tb]
\vspace*{-.3in}
\centerline{\psfig{figure=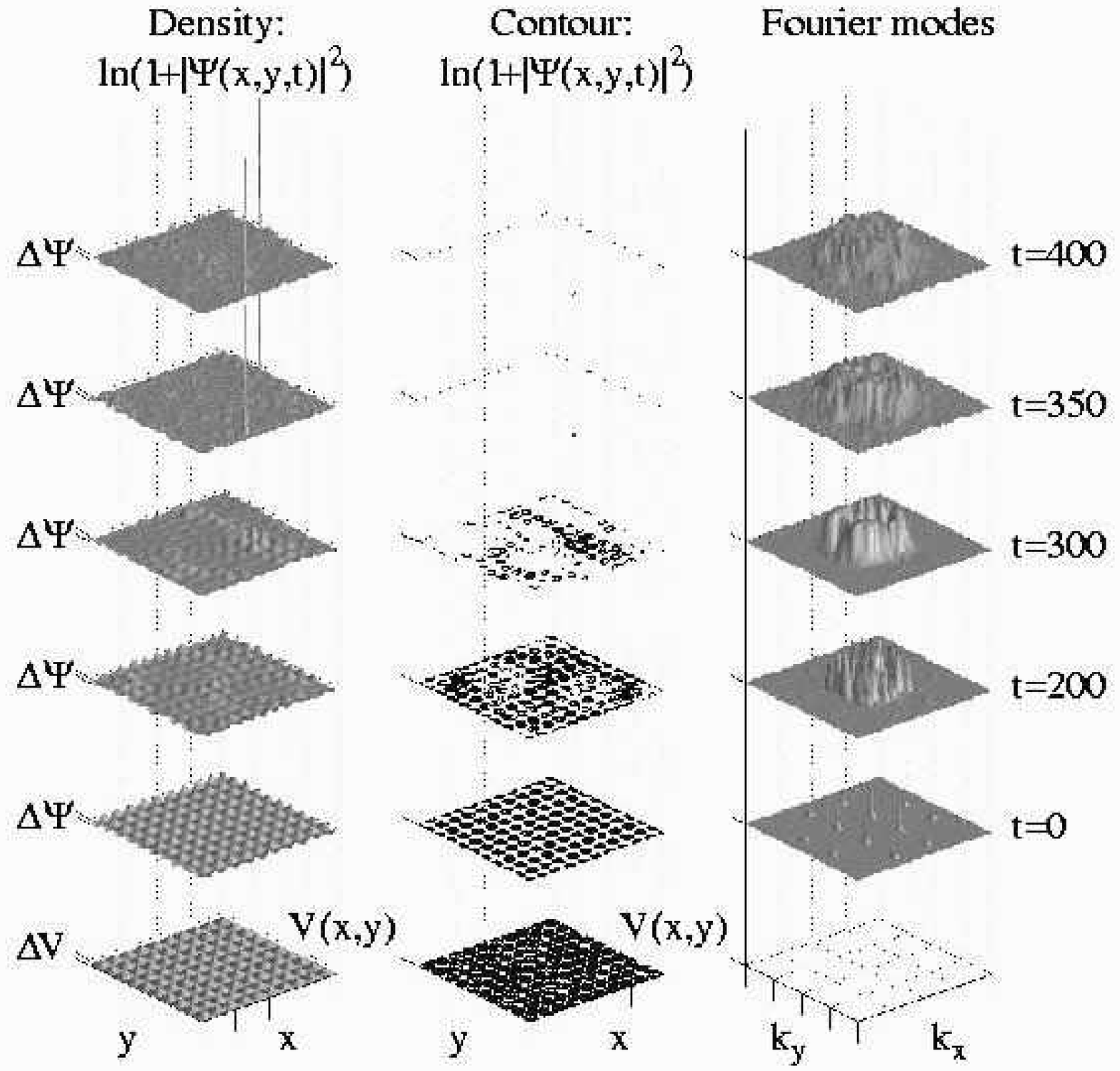,width=130mm,silent=}}
\vspace*{-.4in}
\caption{\label{fig:sfoc-cnstack}
 The two-dimensional unstable attractive evolution of
 the  exact $\cn(m_i x_i,k_i)$ solution corresponding to \rf{eqn:trivphasec} with 16
 periods, and $k_1=k_2=0.5$, $m_1=m_2=1$, and $A_1=A_2=-0.5$.  Note that
 the collapse time is well beyond $t\approx 200$.}
\end{figure}

The $\cn(m_i x_i,k_i)$ solution can be either peak-on-peak or peak-on-trough,
depending on the parameters.  In the absence of the mean-field nonlinearity,
the peak-on-trough  $\cn(m_i x_i,k_i)$ solution would be stable.  However, the
cubic nonlinearity once again gives rise to collapse and blow-up.  The
evolution of a double-amplitude initial condition 
is shown in Fig. \ref{fig:cn-cn-stack} with
$A_1=A_2=-1$.  The collapse of this evolved initial condition 
occurs shortly after $t\approx
0.47$, with the formation of well-defined growing peaks at the locations of the
maxima  of the initial condition.  Since it is the nonlinearity which drives
the collapse, we inhibit the blow-up by choosing an exact solution with
a lower amplitude and smaller resulting nonlinearity.  With $A_1=A_2=-0.5$,
the dynamics of Fig.~\ref{fig:sfoc-cnstack} is 
significantly different from that of Fig. \ref{fig:cn-cn-stack}.  In
particular, the onset of collapse does not occur until $t\approx 200$. 
Furthermore, the $L^2$ norm for the 16 periods considered computationally is
just large enough to
lead to collapse. Thus for small-amplitude solutions, it
may be possible to obtain a periodic condensate in the lattice potential 
over the lifetime of the experiment. Note from Table \ref{table}, that $\delta
t=200$ far exceeds the time scale of any experiment with attractive condensates. 

%
%

%
%

\begin{figure}[tb]
\vspace*{-.2in}
\centerline{\psfig{figure=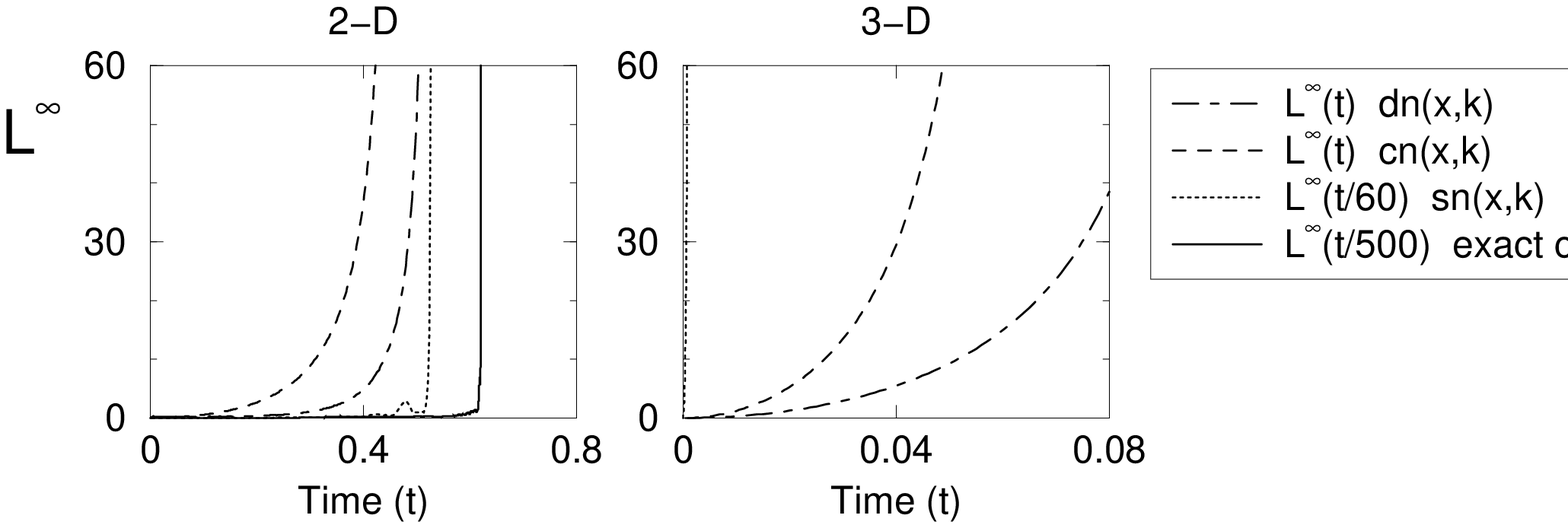,width=120mm,silent=}}
\vspace*{-.1in}
\caption{\label{fig:Lfocus} Evolution of the $L^\infty$ norm of the error 
   in two- 
   and three-dimensions for several trivial-phase solutions of the attractive
   equation.}
\end{figure}

\sloppypar The unstable behavior is further illustrated by the evolution of
the $L^\infty$ norm of the error: ${\rm max}_{\{x,y\}}\left|
  |\psi(x,y,t)|^2-|\psi(x,y,0)|^2 \right|$.  In Fig.  \ref{fig:Lfocus}, the
  dynamics of the 
$L^\infty$ norm is given for both the two-dimensional and
three-dimensional solutions of the attractive equation considered numerically in
this paper.  The $L^\infty$ norm of the unstable attractive solutions grow to
infinity at the onset of the collapse of the peaks.  This
illustrates the fundamental nature of the attractive collapse and blow-up, 
{\em i.e.}, large gradients or sharp peaks develop
in the solution.

\subsubsection{Three-Dimensional Solutions}

\begin{figure}
\vspace*{-.3in}
\centerline{\psfig{figure=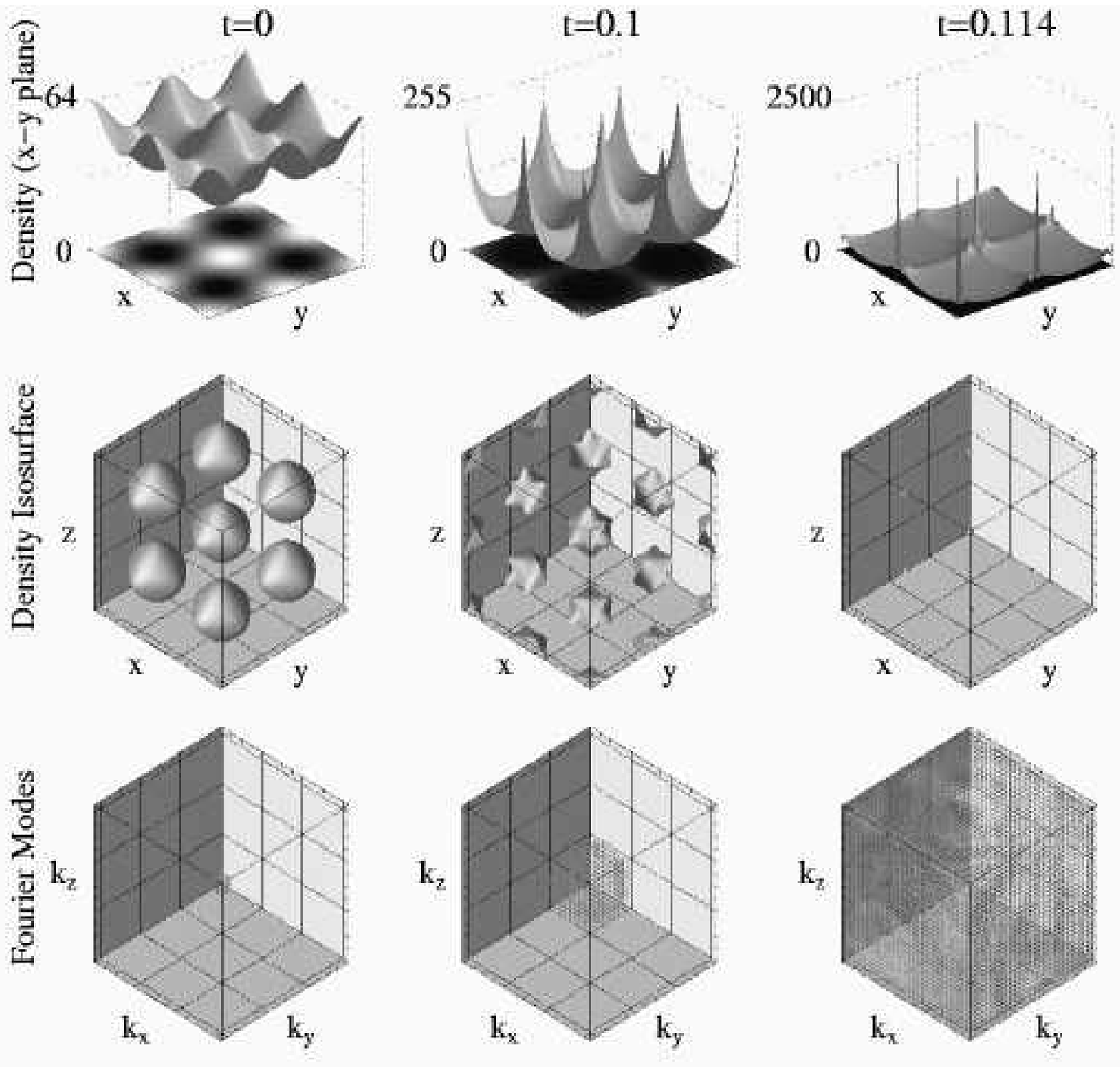,width=100mm,silent=}}
\vspace*{-.2in}
\caption{\label{fig:dnfocus}
 The three-dimensional unstable attractive evolution of
 the double-amplitude $\dn(m_i x_i,k_i)$ initial condition 
 corresponding to \rf{eqn:trivphasec} with
 eight
 periods (two in each direction), 
 with $k_1=k_2=k_3=0.5$, $m_1=m_2=m_3=1$, and $A_1=A_2=A_3=-0.25$.}
\end{figure}

Following the discussion of the two-dimensional solutions, we consider the
trivial phase $\dn(m_i x_i,k_i)$ solution \rf{eqn:trivphasec}.  Also in three
dimensions the plane-wave solution is modulationally unstable and the solution
is linearly unstable.  The unstable behavior however is dominated by collapse
and blow-up which are characteristic of the attractive regime.  The evolution
of two periods of this solution with $A_i=-0.25$, $m_i=1$ and $k=0.5$,
$i=1,2,3$ for $t\in[0,0.114]$ is shown in Fig. \ref{fig:dnfocus}.  As with the
two-dimensional solutions, we expedite the collapse and blow-up by doubling
the amplitude of some of our exact solutions.  The three rows of this figure,
and of all other figures of this type, represent a slice in the $(x,y)$-plane of
the evolution of the condensate density $|\psi(x,y,0,t)|^2$, an iso-surface of
$|\psi(x,y,z,t)|^2$, and the evolution of an iso-surface of the arctan of the
Fourier spectrum.  For the attractive equation, the displayed iso-surface for
the density is at $50\%$ of the value from minimum to maximum while the
iso-surface of the Fourier spectrum is shown at a value of one. 
\hfill All other

%
%

\clearpage

%
%

%
\begin{figure}
\vspace*{-.3in}
\centerline{\psfig{figure=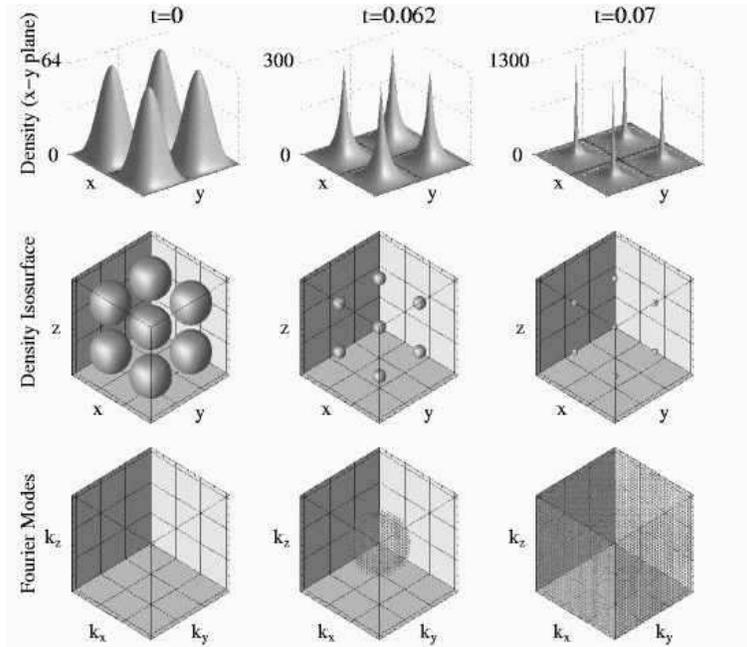,width=100mm,silent=}}
\vspace*{-.2in}
\caption{\label{fig:snfocus}
 The three-dimensional unstable attractive evolution of
 a double-amplitude $\sn(m_i x_i,k_i)$ initial condition 
 corresponding to \rf{eqn:trivphasec} with eight
 periods (two in each direction), 
 with $k_1=k_2=k_3=0.5$, $m_1=m_2=m_3=1$, and $A_1=A_2=A_3=1$.}
\end{figure}
%

\clearpage

\noindent
figures of this type are similar.  This figure clearly shows the localization
of the collapsing solution which results in the spreading of the Fourier modes
in the spectrum.

A double-amplitude $\sn(m_i x_i,k_i)$ peak-on-peak initial condition is also
unstable in the attractive limit as shown in Fig. \ref{fig:snfocus}.  As with the
$\dn(m_i x_i,k_i)$ initial condition,  the dynamics eventually lead to blow-up
near $t\approx 0.07$.  In this case, the total $L^2$ norm is much larger than
for its two-dimensional counterpart so that more distinct peaks are observed
during the collapse process. 

\begin{figure}[tb]
\vspace*{-.3in}
\centerline{\psfig{figure=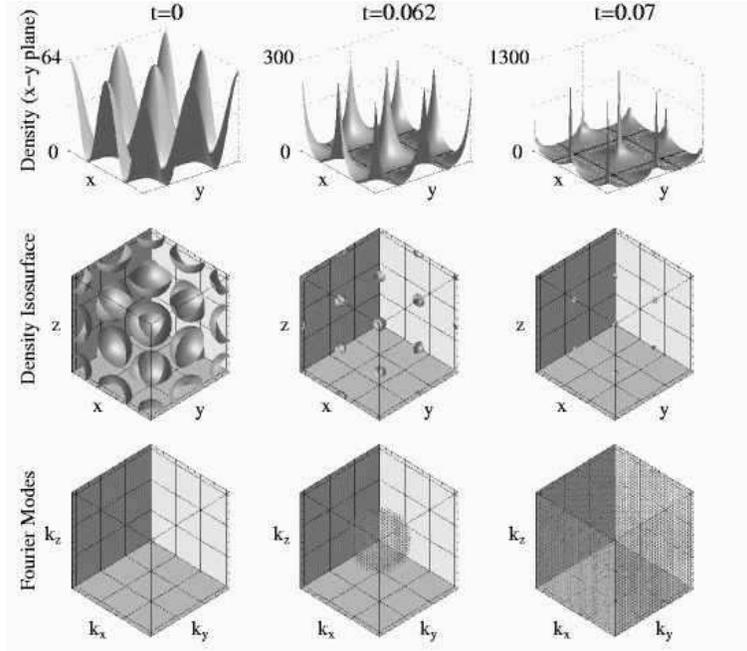,width=100mm,silent=}}
\vspace*{-.2in}
\caption{\label{fig:cnfocus}
 The three-dimensional unstable attractive evolution of
 a double-amplitude $\cn(m_i x_i,k_i)$ initial condition corresponding to 
 \rf{eqn:trivphasec} with eight 
 periods (two in each direction), 
 with $k_1=k_2=k_3=0.5$, $m_1=m_2=m_3=1$, and $A_1=A_2=A_3=-1$.}
\end{figure}

\begin{figure}[tb]
\vspace*{-.3in}
\centerline{\psfig{figure=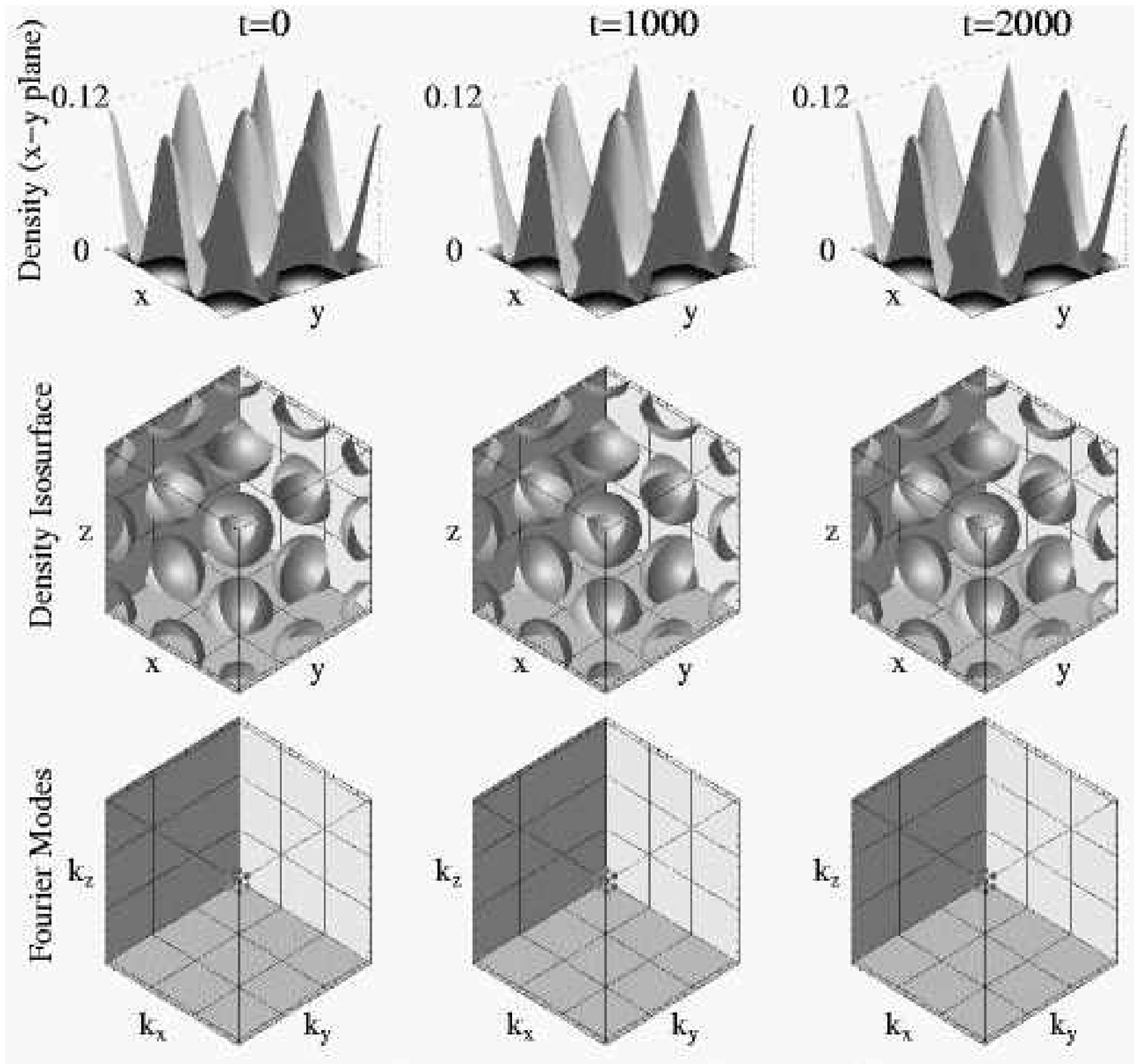,width=100mm,silent=}}
\vspace*{-.2in}
\caption{\label{fig:cnstablefocus}
 The three-dimensional stable attractive evolution of
 the exact $\cn(m_i x_i,k_i)$ solution corresponding to \rf{eqn:trivphasec} 
 with eight
 periods (two in each direction),
 with $k_1=k_2=k_3=0.5$, $m_1=m_2=m_3=1$, and $A_1=A_2=A_3=-0.5$.}
\end{figure}

As in two dimensions, an exact peak-on-trough $\cn(m_i x_i,k_i)$ solution is
stable in the absence of the mean-field nonlinearity. 
However, the cubic nonlinearity once again gives rise to collapse and blow-up.
The evolution of a double-amplitude initial condition is shown in 
Fig. \ref{fig:cnfocus} with
$A_1=A_2=-1$.  The dynamics leads to collapse of this initial condition
at $t\approx 0.07$, and
the formation of a few well-defined peaks is observed.  Since it is the
nonlinearity which drives the collapse phenomena, we inhibit the blow-up
by choosing an exact initial condition with lower amplitude and smaller resulting
nonlinearity.  Choosing $A_1=A_2=-0.5$, the resulting dynamics of Fig.
\ref{fig:cnstablefocus} is shown to be significantly different from those
considered previously.  In particular, the onset of collapse is arrested by
the fact that only two periods in each direction are considered 
so that the $L^2$ norm is small, and the solution remains in the linear regime
for $t\in[0,2000]$.  This again suggests the possibility of
obtaining a stable periodic condensate in a lattice
potential over the lifetime of an experiment.

\section{Discussion}
   
We considered the repulsive and attractive nonlinear Schr\"odinger equation
with an elliptic function potential in two and three dimensions.  Periodic
solutions of this equation were constructed and their dynamical stability was 
investigated analytically and numerically.  
For condensates with repulsive atomic interactions, all stable, trivial-phase
solutions are deformations of the ground state of the linear Schr\"odinger
equation.  These solutions are off-set from the zero level.  This is confirmed
with extensive numerical simulations on the governing nonlinear equation.
Likewise, nontrivial-phase solutions are stable if they are sufficiently
off-set.  Thus for condensates with repulsive interactions, a large number of
condensed atoms is sufficient to form a stable, periodic condensate, as in the
one-dimensional case \cite{becpre1}.  Physically, this implies stability of
states near the Thomas--Fermi limit.

For condensates with attractive atomic interactions, no
stable solutions are found, in contrast to the one-dimensional case
\cite{becpre2}.  However, the time scale for the onset of instability for some
of our solutions far exceeds the time scale of current physical experiments
with attractive condensates.  This occurs when the solutions are localized in
the troughs of the potential and have nodes.  These solutions may be
observable in experiments, given the time scale over which the instability
develops.

Many issues remain open.  In particular, for the majority of our solutions, we
have no analytical results concerning stability or instability.  Nontrivial
phase solutions especially have eluded analysis.  A more detailed
understanding of the collapse and blow up in periodic potentials is necessary.
Of particular interest is the full class of initial conditions which lead to
collapse and blow-up~\cite{rose,weinstein}.  Furthermore, the effect of a
confining potential in addition to a periodic potential warrants
investigation.  These are only a sampling of the many mathematical and
physical issues which arise in this interesting problem.

%
%

%
%

\clearpage

\section*{Acknowledgments} 
The authors acknowledge 
useful conversations with J. C. Bronski and K. Promislow about
the linear stability analysis and with W. P. Reinhardt concerning
Table 1.

\section*{Appendix A:  The ground state of the Nonlinear Schr\"odinger equation
with repulsive atomic interaction}

This appendix is due to Jared C. Bronski. 

\begin{theo}\la{theo3}
A stationary solution $\psi(\vec{x},t)=r(\vec{x})  e^{i(\theta(\vec{x})-\omega
t)}$ with $r(\vec{x})>0$ of  \rf{eqn:nls}, $\alpha=1$  is a global
minimizer of the Hamiltonian \rf{eqn:hamiltonian} with the constraint that
$||\psi(x,t)||^2=C$, for some constant $C$. 
\end{theo}

\no {\bf Proof}~~~
Consider the reduced Hamiltonian
\beq\la{eqn:reduced}
{\hat {\cal H}}(\psi)={\cal H}(\psi)-\lambda ||\psi||^2=\int_{\Omega}
\left(
\frac{1}{2}\nabla \psi\cdot \nabla \psi^*+V(\vec{x}) |\psi|^2+
\frac{1}{2}|\psi|^4-\lambda |\psi|^2
\right)
d\vec{x},
\eeq
where $\lambda$ is a Lagrange multiplier. Its critical points are determined by 
\beq\la{eqn:critical}
\delta {\hat {\cal H}}(\psi)=0 \Rightarrow 
\lambda \psi=-\frac{1}{2} \Delta \psi+|\psi|^2 \psi+V(\vec{x}) \psi.
\eeq
Thus the Lagrange multiplier $\lambda$, which enforces the constraint
of fixed particle number, is interpreted physically as the chemical potential
$\omega$. 
In what follows, the factor $e^{-i \omega t}$ in the
solution is omitted. 
In general, an expression for $\lambda=\omega$ in terms of
the critical point $\psi$ is obtained by multiplying this expression by $\psi^*$
and integrating over $\Omega$:
\beq\la{multiplier}
\omega=\frac{\int_{\Omega}
\left(
\frac{1}{2}\nabla \psi\cdot \nabla \psi^*+V(\vec{x}) |\psi|^2+|\psi|^4
\right)
 d\vec{x}}{||\psi||^2}.
\eeq
Global minimizers of ${\cal H}(\psi)$ subject to the constraint $||\psi||^2=C$
may be assumed to be real, since

\alpheqn
\bea
||r(\vec{x}) e^{i \theta(\vec{x})}||^2&=&||r(\vec{x})||^2\\
{\cal H}(r(\vec{x}) e^{i \theta(\vec{x})})&=&{\cal H}(r(\vec{x}))+\frac{1}{2} 
\int_{\Omega} r^2 (\nabla \theta)^2 d\vec{x}\geq {\cal H}(r(\vec{x})),
\eea
\resetalpheqn
\no with equality only valid for $\nabla \theta=0$. In this case, because of the
phase invariance of \rf{eqn:nls}, we can choose $\theta=0$. Clearly a global
minimizer is only fixed up to multiplication by a constant phase factor.  

The second variation of the reduced Hamiltonian is given by the quadratic form 
\bea\nonumber
\left<\phi \left| \delta^2 {\hat {\cal H}}(\psi) \right|
  \phi\right>\!\!&=&\!\! \int_{\Omega} 
\left(
\frac{1}{2}\nabla \phi \cdot \nabla \phi^*+V(\vec{x})|\phi|^2+\frac{1}{2}\psi^2
{\phi^*}^2+\frac{1}{2}{\psi^*}^2 \phi^2+2 |\psi|^2 |\phi|^2-\omega |\phi|^2
\right)
d\vec{x}\\
\!\!&=&\!\!\left<u|L_+|u\right>+\left<v|L_-|v\right>,
\eea
where $\psi$ is assumed to be real, $\phi=u+i v$, and $L_+$ and $L_-$ are
defined by (\ref{eqn:lplus}-b) with $\theta=0$. From the proof of Theorem
\ref{theo1}, it follows that for $\psi(\vec{x})=r(\vec{x})>0$, the operator
$L_+$ is positive definite, whereas $L_-$ is positive semi-definite. Thus 
$\psi(\vec{x})=r(\vec{x})>0$ is a local minimizer of the Hamiltonian, subject to
the constraint $||\psi||^2=C$. 

The proof that a positive solution is necessarily a global minimizer is
by contradiction. Assume there exists a 
${\hat \psi}$ such that $||{\hat \psi}||^2=C$ and ${\cal H}({\hat \psi}) < 
{\cal H}(\psi)$. By assumption, $\psi$ is real. This can also be assumed for
${\hat \psi}$ since this can only lower the energy. Now consider 
\beq\la{eqn:takethat}
\Psi(\nu)=\cos(\nu) \psi +i \sin(\nu) {\hat \psi}.
\eeq
Then $\Psi(\nu)$ satisfies the constraint $||\Psi(\nu)||^2=C$, since
\beq
||\Psi(\nu)||^2=\cos^2(\nu) ||\psi||^2+\sin^2(\nu) ||{\hat \psi}||^2=C. 
\eeq
Also, 
\bea\nonumber
{\cal H}(\Psi(\nu))&=&\cos^2(\nu) \int_\Omega \left( 
\frac{1}{2}\nabla \psi \cdot \nabla \psi^*+V(\vec{x})
|\psi|^2+\frac{1}{2}\cos^2(\nu)|\psi|^4
\right) d\vec{x}+\\\nonumber
&&\sin^2(\nu) \int_\Omega \left( 
\frac{1}{2}\nabla {\hat \psi} \cdot \nabla {\hat \psi}^*+V(\vec{x})
|{\hat \psi}|^2+\frac{1}{2}\sin^2(\nu)|{\hat \psi}|^4
\right) d\vec{x}+\\\nonumber
&&\cos^2(\nu) \sin^2(\nu)\int_\Omega |\psi|^2 |{\hat
\psi}|^2d\vec{x}\\\la{convexity}
&\leq& \cos^2(\nu) {\cal H}(\psi)+\sin^2(\nu) {\cal H}({\hat \psi}).
\eea
This last inequality follows from $\int_{\Omega}\left(\cos^4(\nu)|\psi|^4+2
\cos^2(\nu)\sin^2(\nu)|\psi|^2|{\hat \psi}|^2+
\sin^4(\nu)|{\hat \psi}|^4\right)d\vec{x}\leq
\int_{\Omega}\left(\cos^2(\nu)|\psi|^4+
\sin^2(\nu)|{\hat \psi}|^4\right)d\vec{x}$, which is easily verified. Equation
\rf{convexity} expresses the convexity of the Hamiltonian as a functional of
$\psi$. Now, since ${\cal H}({\hat \psi}) < {\cal H}(\psi)$, near $\nu=0$ 
\bea\nonumber
{\cal H}(\Psi(\nu))&\leq& \left(1-\frac{\nu^2}{2}+{\cal O}(\nu^4)\right)^2 
{\cal H}(\psi)+\left(\nu-\frac{\nu^3}{6}+{\cal O}(\nu^5)\right)^2 
{\cal H}({\hat \psi})\\\la{quadratically}
&\leq& {\cal H}(\psi)-\nu^2 \left(
{\cal H}(\psi)-{\cal H}({\hat \psi})
\right)+{\cal O}(\nu^4),
\eea
and thus ${\cal H}(\Psi(\nu))$ is quadratically decreasing near $\nu=0$. But,
\bea\nonumber
\frac{1}{2}\left.\frac{d^2}{d \nu^2}{\cal H}(\Psi(\nu))\right|_{\nu=0}&=&
- \int_\Omega \left( 
\frac{1}{2}\nabla \psi \cdot \nabla \psi^*+V(\vec{x})
|\psi|^2+|\psi|^4
\right) d\vec{x}+\\\nonumber
&& \int_\Omega \left( 
\frac{1}{2}\nabla {\hat \psi} \cdot \nabla {\hat \psi}^*+V(\vec{x})
|{\hat \psi}|^2+|\psi|^2 |{\hat
\psi}|^2
\right) d\vec{x}\\\nonumber
&=&\int_\Omega \left( 
\frac{1}{2}\nabla {\hat \psi} \cdot \nabla {\hat \psi}^*+V(\vec{x})
|{\hat \psi}|^2+|\psi|^2 |{\hat
\psi}|^2
\right) d\vec{x}-\omega \int_\Omega |\psi|^2 d\vec{x}\\
&=&\left<{\hat \psi}\Big|L_-\Big|{\hat \psi}\right>\geq 0,
\eea
where we have used \rf{multiplier}. This contradicts the quadratic decrease of 
${\cal H}(\Psi(\nu))$ near $\nu=0$. Thus, no ${\hat \psi}$ exists which
satisfies $||{\hat \psi}||^2=C$ and which has a lower energy value than $\psi$.
This proves the theorem. \hspace*{\fill}$\bbox$

\section*{Appendix B:  Collapse and blow-up} 

A comprehensive overview of many aspects of collapse and blow-up for the
attractive ($\alpha=-1$) nonlinear Schr\"odinger equation \rf{eqn:nls}  with
$V(\vec{x})=0$ is given in \cite{sulem}. However, some modification of the
theory is required for nonzero potential $V(\vec{x})\neq 0$. In
\cite{pitaevskii2,wadati} such modifications were considered for harmonic trap
potentials. However, part of the analysis given in \cite{wadati} applies to
general potentials. This is the starting point of our considerations. 

For the remainder of this appendix, we consider condensates with attractive
atomic interactions for which $\alpha=-1$. Equation \rf{eqn:nls} has two conserved
quantities, given by the $L^2$-norm of the solution (this has an interpretation
as the number of particles per period $N$) and the Hamiltonian \rf{eqn:hamiltonian}.
In quantum mechanics, the nonlinear term in \rf{eqn:nls} is absent, and
$\psi(\vec{x},t)$ has the interpretation of a probability density function. This
allows the definition of the expected value of a physical quantity. For
instance, 
\beq\la{expectedx}
\vec{X}=<\vec{x}>=\frac{1}{N}\int_\Omega \vec{x} |\psi|^2 d\vec{x}
\eeq
is the expected value of the position. Such quantities can still be constructed
in the presence of the nonlinearity,
although some care is required to interpret them. The above definition leads to 
\beq\la{newton}
\frac{d^2 \vec{X}}{d t^2}=-\left<\nabla V(\vec{x})\right>. 
\eeq
This equation is Newton's law. In quantum mechanics, 
the expected values of physical quantities
satisfy classical equations of motion. That this equation persists in the
presence of a nonlinear term is surprising. 

To examine collapse, the quantity \cite{sulem}
\beq\la{variance}
W(t)=\int_{\Omega} \left(\sum_{i=1}^d x_i^2\right) |\psi|^2 d\vec{x}
\eeq
is studied. Note that it is by definition a positive quantity. The solution of
\rf{eqn:nls} is said to collapse if $W(t)$ becomes negative \cite{sulem,wadati}.
Use of the conserved quantities and Newton's law \rf{newton} 
gives the variance identities
\alpheqn
\bea\la{var1}
\frac{d^2 W}{d t^2}&=&2 \int_\Omega |\psi|^2 d\vec{x}-d \int_\Omega |\psi|^4
d\vec{x}-2 \int_\Omega |\psi|^2 (\vec{x} \cdot \nabla V(\vec{x})) d\vec{x}\\
\la{varid}
&=&4 {\cal H}(\psi)-(d-2)\int_\Omega |\psi|^4 d\vec{x}-2
\int_\Omega |\psi|^2 (2 V(\vec{x})+\vec{x}\cdot \nabla V(\vec{x})) d\vec{x}
\eea
\resetalpheqn
These identities are generalizations of variance identities for the free
nonlinear Schr\"odinger equation ($V(\vec{x})=0$). With $d\geq 2$,
\beq\la{varineq}
\frac{d^2 W}{d t^2}\leq 4 {\cal H}(\psi)-2
\int_\Omega |\psi|^2 G(\vec{x}) d\vec{x},
\eeq
with $G(\vec{x})=2 V(\vec{x})+\vec{x}\cdot \nabla V(\vec{x})$. For specific
choices of the potential (such as a harmonic trap, for which
$\Omega=\mathbb{R}^d$) a more detailed analysis of the collapse is 
possible. However, for the potential \rf{eqn:pot}, the analysis here
is
restricted to general statements. 

If $G(\vec{x})\geq 0$ then 
\bea\la{finalineq}
&\ds\frac{d^2 W}{d t^2}\leq 4 {\cal H}(\psi)&\\\la{collapsetime}
&\Rightarrow W(t)\leq 2 {\cal H}(\psi) t^2+\beta t+\gamma.& 
\eea
If for the initial condition ${\cal H}(\psi)\leq 0$, then for all $t$, 
${\cal H}(\psi)\leq 0$, since the Hamiltonian is conserved. Then $d^2 W/d
t^2\leq 0$, and for some time $t_c$, $W(t)<0$ if $t>t_c$. A crude overestimate of
this time $t_c$ is found from \rf{collapsetime}. Presumably it is possible for
solutions with positive energy to collapse as well, but this requires more
subtle arguments \cite{wadati,weinstein} than those used here. 

Thus, for a given potential one checks whether $G(\vec{x})\geq 0$. If so, then every
initial condition of \rf{eqn:nls} with negative energy collapses in finite time. Note that
it is always possible to ensure $G(\vec{x})\geq 0$ on $\Omega$:  this can be
achieved by adding a constant $V_0$ to the potential $V(\vec{x})$, since
$\Omega$ is a bounded domain.  However, this increases the energy of the
solution by an amount $N V_0$. Note that the absolute level of the potential is
physically irrelevant, which is clear from \rf{var1}, where only the gradient of
the potential appears. In order to maximize the set of initial conditions for
which collapse can be concluded, the constant $V_0$ is chosen such that
$G(\vec{x})\geq 0$, where equality is achieved in $\Omega$. 

The exact solutions constructed in Section \ref{sec:exact} are stationary, thus
$W(t)$ is constant for these solutions and $d^2 W/d t^2=0$. Thus, no
conclusion for these solutions
follows from our analysis. By increasing the amplitude of 
the initial conditions, we no longer find exact stationary solutions. However,
it is possible to show that these initial conditions lead to finite-time
collapse. This is the approach followed for many of the numerical runs in two
and three dimensions. 

Finally, following \cite{wadati}, it is easy to show that collapse of a solution
implies blow-up of that solution, $i.e.,$ $\sup_\Omega |\psi|\rightarrow
\infty$. The argument here is easier than the argument in \cite{wadati} 
because $\Omega$ is bounded.   

\begin{theo}
For condensates with attractive atomic interactions ($\alpha=-1$),
collapse implies blow-up.
\end{theo}

\no {\bf Proof}~~~
Consider
\beq
\int_\Omega |\psi|^4 d\vec{x}\leq \left(\sup_\Omega |\psi|\right)^2 
\int_\Omega |\psi|^2 d\vec{x}=N \left(\sup_\Omega |\psi|\right)^2. 
\eeq 
The Heisenberg uncertainty principle states 
\beq
W(t)
\int_\Omega  \nabla \psi \cdot \nabla \psi^* d\vec{x}\geq 
\left(\frac{d N}{2}\right)^2,
\eeq
thus 
\bea
\left(\sup_\Omega |\psi|\right)^2&\geq& \frac{1}{N} \int_\Omega |\psi|^4
d\vec{x}
\nonumber\\
&=&\frac{1}{N}\left(
-2 {\cal H}(\psi)+
\int_\Omega  \nabla \psi \cdot \nabla \psi^* d\vec{x}+2 \int_\Omega V(\vec{x})
|\psi|^2 d\vec{x}\right)\nonumber \\
&\geq&\frac{1}{N}\left(
-2 {\cal H}(\psi)+\left(\frac{d N}{2}\right)^2\frac{1}{W}
+2 \int_\Omega V(\vec{x})|\psi|^2 d\vec{x}
\right).
\eea
Since $\int_\Omega V(\vec{x})|\psi|^2 d\vec{x}\in [-N \sup_\Omega V(\vec{x}), 
N \sup_\Omega V(\vec{x})]$, this implies that as $W(t)\rightarrow 0$,
$\sup_\Omega |\psi|\rightarrow \infty$. 
\hspace*{\fill}$\bbox$



\end{document}